\documentclass[a4paper,oneside,final,notitlepage,onecolumn,12pt]{article}
\usepackage{amsfonts}
\usepackage{epsf}
\usepackage{graphicx}% Include figure files
\usepackage{amssymb,eso-pic}
\usepackage{latexsym}
\usepackage{tabularx}
\usepackage{amsxtra} 
\usepackage{hyperref}
\usepackage{t1enc}
\usepackage{amsmath,accents}
\usepackage{bbm}
\usepackage{enumerate}
\usepackage{cancel}

\usepackage{ifpdf,epsfig,array,amsmath,amssymb}

\usepackage{psfrag} 
\psfrag{na}{\footnotesize $n^a$}   
\psfrag{ta}{\footnotesize $\sigma^a$}   
\psfrag{vna}{\footnotesize $\check{n}^a$}
\psfrag{vNa}{\footnotesize $\check{N}^a$}
\psfrag{scrW}{\footnotesize $\mycal{W}_{\rho_0}$}
\psfrag{t=t1}{\footnotesize $\Sigma_{\sigma_1}$}
\psfrag{t=t2}{\footnotesize $\Sigma_{\sigma_2}$}
\psfrag{r=all}{\footnotesize $\rho=const$}
\psfrag{hna}{\footnotesize $\,\widehat{n}^a$}
\psfrag{tna}{\footnotesize $\tilde{n}^a$}
\psfrag{hab,Kab}{\footnotesize $h_{ab}, K_{ab}$}
\psfrag{thab,tKab}{\footnotesize $\tilde h_{ab}, \tilde K_{ab}$}
\psfrag{hhab,hKab}{\footnotesize $\widehat h_{ab}, \widehat K_{ab}$}
\psfrag{chab,cKab}{\footnotesize $\check{h}_{ab}, \check{K}_{ab}$}

\usepackage[normalem]{ulem}
\usepackage{color}
\definecolor{blue}{rgb}{0,0,1}
\definecolor{red}{rgb}{1,0,0}

\definecolor{DGREEN}{rgb}{0,0.7,0.3}
\definecolor{grey1}{rgb}{0.52, 0.52, 0.51}

\newcommand{\interior}[1]{\accentset{\smash{\raisebox{-0.1ex}{$\scriptstyle\circ$}}}{#1}\rule{0pt}{2.3ex}}
\newcommand{\instar}[1]{\accentset{\smash{\raisebox{-0.12ex}{$\scriptstyle\star$}}}{#1}\rule{0pt}{2.3ex}}
\newcounter{mnotecount}%[section]

\newcommand{\mnotex}[1]%{}
{\protect{\stepcounter{mnotecount}}$^{\mbox{\footnotesize $\bullet$\themnotecount}}$ 
	\marginpar{%\color{red}%
		\raggedright\tiny\em
		$\!\!\!\!\!\!\,\bullet$\themnotecount: #1} }


\makeatletter
\def\@xfootnote[#1]{%
  \protected@xdef\@thefnmark{#1}%
  \@footnotemark\@footnotetext}
\makeatother

\DeclareFontFamily{OT1}{rsfs}{} \DeclareFontShape{OT1}{rsfs}{m}{n}{
<-7> rsfs5 <7-10> rsfs7 <10-> rsfs10}{}
\DeclareMathAlphabet{\mycal}{OT1}{rsfs}{m}{n}

\def\sc{{\hskip 3.5pt {{}^{{}^{{}_{{}_{\bowtie}}}}} \kern -8.pt{}}}  
\def\SC{{\hskip 3.5pt {{}^{{}^{{}^{{}_{{}_{\bowtie}}}}}} \kern -10.5pt{}}}

\def\d{{\rm d}}

\DeclareMathAlphabet{\mathpzc}{OT1}{pzc}{m}{it}

\newcommand{\hoch}[1]{$\, ^{#1}$}

\newcommand{\auth}{  Istv\'an R\'{a}cz\hoch{1,3}
and \ Jeffrey Winicour\hoch{2,3}
}

\begin{document}

%%%%%%%%%%%%%%%%%%%%%%%%%%%%%%%%%%%%%%%%%%%%%%
\newtheorem{theorem}{Theorem}[section]
\newtheorem{lemma}{Lemma}[section]
\newtheorem{proposition}{Proposition}[section]
\newtheorem{corollary}{Corollary}[section]
\newtheorem{conjecture}{Conjecture}[section]
\newtheorem{example}{Example}[section]
\newtheorem{definition}{Definition}[section]
\newtheorem{remark}{Remark}[section]
\newtheorem{exercise}{Exercise}[section]
\newtheorem{axiom}{Axiom}[section]
%%%%%%%%%%%%%%%%%%%%%%%%%%%%%%%%%%%%%%%%%%%%%
\renewcommand{\theequation}{\thesection.\arabic{equation}}

\begin{center}

{\LARGE{\bf Toward computing gravitational initial
data without elliptic solvers} }

\vspace{25pt}
\auth

\vspace{10pt}{\hoch{1}\it Wigner Research Center for Physics} \\  {H-1121 Budapest, Hungary}

\vspace{10pt}{\hoch{2}\it Department of Physics and Astronomy,\\ 
University of Pittsburg, Pittsburgh, PA, 15260,  USA}

\vspace{10pt}{\hoch{3}\it Max Planck Institute for Gravitational Physics} \\ 
 {Albert Einstein Institute, Golm, Germany}

\vspace{15pt}

\today
%\date{July 4, 2015}

\begin{abstract}

Two new methods have been proposed for solving
the gravitational constraints without using elliptic solvers
by formulating them as either an algebraic-hyperbolic
or parabolic-hyperbolic system. Here, we compare
these two methods and present a unified
computational infrastructure for their
implementation as numerical evolution codes.
An important potential application of these methods
is the prescription of initial data for the
simulation of black holes. This paper is meant to support progress and
activity in that direction.

\end{abstract} 

\end{center}

\section{Introduction}\label{introduction}
\setcounter{equation}{0}

Physically realistic initial data are of
major importance for the
numerical simulation of gravitational systems such as binary black holes.
The prescription of the initial data is complicated mathematically
by the nonlinear constraint equations that they must satisfy. Traditionally,
the constraints have been formulated
as elliptic equations,
based upon the conformal treatment of the Hamiltonian constraint
by Lichnerowicz~\cite{lich}
and later extended by York~\cite{york0,york1} to treat the momentum
constraint. For reviews see~\cite{cook,gourg}.

Recently, two alternative methods for solving the constraints by means
of evolution systems
were introduced in \cite{racz_constraints} 
(see also  \cite{racz_geom_det,racz_geom_cauchy,racz_tdfd}). 
In one of the proposed methods the Hamiltonian
constraint is solved algebraically and the 
momentum constraints are expressed as a first order symmetric hyperbolic system.
In the other method,
the Hamiltonian constraint is formulated as a
parabolic equation, with the
momentum constraints again expressed as a symmetric hyperbolic system.
Both of these two methods of solving the constraints
have been shown to lead to well-posed problems
for the fully nonlinear systems. Note that a well-posed problem is a necessity
for a stable numerical simulation.
In particular, as an important first step in establishing the viability of the
algebraic-hyperbolic method,
it was shown that a condition necessary for numerical stability holds globally
in the case of nonlinear perturbations of Schwarzschild black hole data \cite{i_jeff}.

The details of the waveform supplied by numerical simulation of the inspiral
and merger of  a binary black hole
is key input for interpreting the scientific content of the signals which have been observed
by the LIGO-Virgo collaboration. Thus it is important that the initial data does
not introduce spurious effects, e.g. the  initial ``junk radiation''
common to all simulations based
upon elliptic formulations of the constraint problem.
Elliptic equations require boundary data at inner boundaries
in the strong field region
near the singularities inside the black holes, as well as at an outer boundary
in the far field. Neglect of the tidal interaction beteween the black holes in a binary is a
likely source of the junk radiation~\cite{chu}.  Other sources of junk radiation have been
traced to the use of conformally flat iniial
data. However, alternatives to conformal flatness
have reduced the junk radiation content by only a factor
of order 2.~\cite{chu,lovelace}.  This complicates
the important problem of matching a numerical evolution to post-Netwonian parameters.
Currently, this matching must be done after the junk radiation subsides.

Here we present the calculational details of two methods to solve the constraints which do not
involve elliptic equations and only require
data on the outer boundary, where the choice of
boundary data can be guided by asymptotic flatness.
In the algebraic-hyperbolic system, the 3-metric of the
initial hypersurface and
two components of external curvature
corresponding to the radiative degrees of freedom
are prescribed freely. The remaining components of external curvature 
are determined from the constraints. For binary black hole data, the
3-metric data can prescribed in superposed  Kerr-Schild
form for the individual black holes, as in~\cite{ksm1,ksm2}.
The two components of extrinsic curvature data representing the gravitational degrees
of freedom can also be  prescribed by superposing the individual black hole data.
In a linear theory, the superposition of such initial data for a non-radiative solution
would lead to a non-radiative solution.This provides encouragement
that this method might suppress junk radiation.  A similar strategy is possible
for the parabolic-hyperbolic system~\cite{racz_ADM}. However, due to the
nonlinearity of the constraints, there is no guarantee that, in the strong field
region near the black holes, the constrained
components of extrinsic curvature would not introduce spurious radiation. This is an issue for future work.

The constraints are solved by means
of an  inward ``evolution'' from the
outer boundary by either a parabolic-hyperbolic
or an algebraic-hyperbolic system of equations. Numerical
stability has been demonstrated in simulations
of initial data for a single perturbed black hole by means of both the algebraic-hyperbolic
system~\cite{babac} and the parabolic-hyperbolic
system~\cite{ALI}. Boundary data are only necessary at
the outer boundary, where their choice can be guided by
asymptotic flatness.

Brief technical notes presenting a pseudo-code for the numerical solution of the
algebraic-hyperbolic system were posted earlier \cite{i_jeff_2}. Since then, there has
been activity in implementing both the algebraic-hyperbolic constraint
system \cite{beyer2,babac,maciej} and the parabolic-hyperbolic system \cite{Christian,ALI}.
Because of this interest, and what we consider to be the
importance of the problem, here we extend the scope of
these technical notes in the following
two directions. First, we present a unified treatment of the
computational infrastructure necessary to implement the
two approaches as numerical evolution codes.
Second, we describe the details of the foliation, lapse and
shift necessary for the formulation of a Cauchy problem
for constructing the initial data. Our purpose is to supply the
computational infrastructure for further code development and exploration.

In Sec.~\ref{preliminaries}, we review 
the main ideas behind these two methods.
In both methods, the choice of foliation of the initial Cauchy hypersurface plays
an important role. 

In Sec.~\ref{sec:foliations}, we discuss two simple choices
of foliation, by spheres or by planes, for integrating the resulting constraint systems.
In Sec.~\ref{decomps}, we describe how to
decompose the basic fields and their derivatives in terms
of the background unit sphere geometry in the case of a spherical
foliation, or a background Euclidean geometry for a 
planar foliation. Finally, in Sec.~\ref{sec:newvar},
we present the explicit form of the constraint systems in terms of spin-weighted
fields defined with respect to the background geometries.

Numerical implementation of the hyperbolic equations is
flexible since the evolution can proceed locally
and is reversible.
Parabolic equations, like elliptic equations, have strong smoothing properties and
the condition for numerical stability can be relaxed  by applying an implicit scheme.
The precision and cost achieved in computing parabolic-hyperbolic initial data in~\cite{ALI} is  comparable to the elliptic approach, as no iteration is necessary. 

%%%%%%%%%%%%%%%%

\section{Preliminaries}\label{preliminaries}
\setcounter{equation}{0}

\medskip

In general relativity the vacuum initial data 
on a three-dimensional manifold $\Sigma$
consist of a Riemannian metric $h_{ij}$ and a symmetric tensor
field $K_{ij}$ (the extrinsic curvature of $\Sigma$).
The pair $(h_{ij},K_{ij})$ is said to satisfy the vacuum constraints
(see e.g.~Refs.~\cite{choquet,wald}) if the relations  
\begin{align} 
{}^{{}^{(3)}}\hskip-1mm R + \left({K^{j}}_{j}\right)^2 - K_{ij} K^{ij} =0\,, 
\label{new_expl_eh}\\
D_j {K^{j}}_{i} - D_i {K^{j}}_{j} =0\label{new_expl_em}%\,,
\end{align}
hold on $\Sigma$, where ${}^{{}^{(3)}}\hskip-1mm R$ and $D_i$
denote the scalar curvature and the
covariant derivative operator associated with $h_{ij}$, respectively. 

The algebraic-hyperbolic and parabolic-hyperbolic constraint systems both introduce
a foliation of  $\Sigma$ by a one-parameter family of two-surfaces $\mycal{S}_\rho$
which are the $\rho=const$ surfaces of a smooth function $\rho$ with
non-vanishing gradient. The constraint system is solved by an
evolution along the streamlines
of a smooth vector field $\rho^i$ on $\Sigma$,
scaled such that $\rho^i \partial_i \rho=1$. Here $\rho^i$
is the analogue of the time evolution vector in a Cauchy
evolution.  

The unit normal $\widehat n^i$
to the $\mycal{S}_\rho$ level surface has the
decomposition
\begin{equation}\label{nhat}
\widehat n^i={\,\widehat{N}}^{-1}\,[\, \rho^i-{\widehat N}{}^i\,]\,,
\end{equation}
where the ``lapse'' $ \widehat N$ and ``shift'' $\widehat N^i$ of the evolution field $\rho^i$
are determined by
$\widehat n_i= \widehat N \partial_i \rho$ and $\widehat N^i=\widehat \gamma{}^i{}_j\,\rho^j$.
Here $\widehat \gamma{}^i{}_j=\delta{}^i{}_j-\widehat n{}^i\widehat n_j$ is the projection
operator corresponding to the normal decomposition of the 
3-metric $h_{ij}$  into the induced metric $\widehat \gamma_{ij}$
of the surfaces $\mycal{S}_\rho$,
\begin{equation}\label{hij}
h_{ij}=\widehat \gamma_{ij}+\widehat  n_i \widehat n_j\,.
\end{equation} 

The extrinsic curvature has the analogous decomposition
\begin{equation}
K_{ij}= \boldsymbol\kappa \,\widehat n_i \widehat n_j  
  + \left[\widehat n_i \,{\rm\bf k}{}_j  
  + \widehat n_j\,{\rm\bf k}{}_i\right]  + {\rm\bf K}_{ij}\,,
\end{equation}
where $\boldsymbol\kappa= \widehat n^k\widehat  n^l\,K_{kl}$,
${\rm\bf k}{}_{i} = {\widehat \gamma}^{k}{}_{i} \,\widehat  n^l\, K_{kl}$
and ${\rm\bf K}_{ij} = {\widehat \gamma}^{k}{}_{i} {\widehat \gamma}^{l}{}_{j}\,K_{kl}$. 
Note that boldfaced symbols, along with the induced metric
$\widehat \gamma_{ij}$  and the shift vector $\widehat N^i$, denote
well-defined fields intrinsic to the 2-surfaces $\mycal{S}_\rho$.
The reformulation of the Hamiltonian and momentum constraints
(\ref{new_expl_eh}) and (\ref{new_expl_em}) also involves
the trace and the trace-free parts of ${\rm\bf K}_{ij}$,
\begin{equation}\label{intK}
{\rm\bf K}^l{}_{l}=\widehat\gamma^{kl}\,{\rm\bf K}_{kl} 
   \quad {\rm and} \quad \interior{\rm\bf K}_{ij}={\rm\bf K}_{ij}
   -\tfrac12\,\widehat \gamma_{ij}\,{\rm\bf K}^l{}_{l} \, .
\end{equation}
In addition, we denote the extrinsic curvature of
 $\mycal{S}_\rho$ by
\begin{equation}\label{hatextcurv}
\widehat K_{ij}= {{\widehat \gamma}^{l}}{}_{i}\, D_l\,\widehat n_j
  =\tfrac12\,\mycal{L}_{\widehat n} {\widehat \gamma}_{ij}
  =\tfrac12\,\widehat{N}^{-1}\left[\mycal{L}_{\rho} {\hat \gamma}_{ij} 
  -\hat  D_{(i}\widehat N_{j)}\right]\,.
\end{equation}

The data pair $(h_{ij},K_{ij})$ can be replaced by the 
above fields 
$\widehat N,\widehat N^i,\widehat \gamma_{ij}, 
\interior{\rm\bf K}_{ij}, \boldsymbol\kappa,{\rm\bf k}{}_{i}$ and ${\rm\bf K}^l{}_{l}$.
It is remarkable that regardless of the choice of foliation or
evolution vector field the Hamiltonian and momentum constraints
(\ref{new_expl_eh}) and (\ref{new_expl_em}) can be re-formulated
as either a parabolic-hyperbolic or algebraic-hyperbolic
evolution system according to the following
constructions, as formulated in \cite{racz_constraints}.

\subsection{The parabolic-hyperbolic constraint system}

In the parabolic-hyperbolic approach, the Hamiltonian constraint is
re-expressed as a parabolic equation (\ref{bern_pde}) for the lapse
$\widehat N$ of the foliation and the momentum constraint is recast
as the first order symmetric hyperbolic system
(\ref{par_const_n})--(\ref{ort_const_n}) for
${\rm\bf k}{}_{i}$ and ${\rm\bf K}^l{}_{l}$, 
\begin{align}
{}& \instar{K}\,[\,(\partial_{\rho} \widehat N) 
- \widehat N{}^l(\hat D_l\widehat N) \,] - \widehat N^{2} (\hat D^l \hat D_l \widehat N)
 - \mathcal{A}\,\widehat N - \mathcal{B}\,\widehat N{}^{3} = 0 \,, \label{bern_pde} \\ 
{}& \mycal{L}_{\hat n} {\rm\bf k}{}_{i} - \tfrac12\,\hat D_i ({\rm\bf K}^l{}_{l}) 
- \hat D_i\boldsymbol\kappa + \hat D^l \interior{\rm\bf K}{}_{li} 
 + \widehat N{}^{-1}\instar{K}\,{\rm\bf k}{}_{i}  
 + [\,\boldsymbol\kappa-\tfrac12\, ({\rm\bf K}^l{}_{l})\,]\,\dot{\hat n}{}_i 
 - \dot{\hat n}{}^l\,\interior{\rm\bf K}_{li} = 0 \label{par_const_n}, \\
{}& \mycal{L}_{\hat n}({\rm\bf K}^l{}_{l}) - \hat D^l {\rm\bf k}_{l} 
- \widehat N{}^{-1}\instar{K}\,[\,\boldsymbol\kappa-\tfrac12\, ({\rm\bf K}^l{}_{l})\,]  
+ \widehat N{}^{-1}\interior{\rm\bf K}{}_{kl}\instar{K}{}^{kl} 
+ 2\,\dot{\hat n}{}^l\, {\rm\bf k}_{l}  = 0\, .
 \label{ort_const_n}
\end{align}
Here $\hat D_i$ stands for the covariant derivative operator associated with
$\hat \gamma_{ij}$ and $\dot{\hat n}{}_k={\hat n}{}^lD_l{\hat n}{}_k
=-{\hat D}_k(\ln{\widehat N})$, and we introduce the notation 
\begin{align}
{}& \instar{K}_{ij}=\tfrac12\mycal{L}_{\rho} {\hat \gamma}_{ij} 
-\hat  D_{(i}\widehat N_{j)}\,, \label{instarK} \\ 
{}&  \instar{K}  =\tfrac12\,{\hat \gamma}^{ij}\mycal{L}_{\rho} {\hat \gamma}_{ij} 
-  \hat D_j\widehat N^j\,,\label{eq:trhatext} \\
{}& \mathcal{A} =(\partial_{\rho} \instar{K}) - \widehat N{}^l (\hat D_l \instar{K}) 
+ \tfrac{1}{2}[\,\instar{K}^2 + \instar{K}{}_{kl} \instar{K}{}^{kl}\,], \label{A}  \\
{}& \mathcal{B} =  -\tfrac12\,\bigl[\widehat{R} + 2\,\boldsymbol\kappa\,({\rm\bf K}^l{}_{l})
  +\tfrac12\,({\rm\bf K}^l{}_{l})^2  -2\,{\rm\bf k}{}^{l}{\rm\bf k}{}_{l} 
   - \interior{\rm\bf K}{}_{kl}\,\interior{\rm\bf K}{}^{kl}\,\bigr] \label{B} \, .
\end{align}

In the form (\ref{bern_pde}), the Hamiltonian constraint is a strongly parabolic
spartial differential equation in the region of $\Sigma$ where $\instar{K}$ is
either strictly positive or strictly negative. In this case,
$\rho$ plays the role of ``time'' and  $\rho^i$ plays the
role of a ``time'' evolution vector field. (For more details see \cite{racz_constraints}).
Note that the sign of $\instar{K}$
determines the stable evolution direction for the parabolic equation.
It is also important that the subsystem (\ref{par_const_n})--(\ref{ort_const_n})
comprises a first order symmetric hyperbolic system. 

As a result, the coupled parabolic--hyperbolic system (\ref{bern_pde})--(\ref{ort_const_n})
possesses a well-posed initial value problem for the dependent variables
$\widehat N, {\rm\bf k}{}_{i}, {\rm\bf K}^l{}_{l}$,
which guarantees the existence of
a local solution.
In solving (\ref{bern_pde})--(\ref{ort_const_n}), the variables
$\widehat N, {\rm\bf k}{}_{i}, {\rm\bf K}^l{}_{l}$ are determined by the constraints
whereas the remaining {four} fields 
$\widehat N^i,\widehat \gamma_{ij}, \boldsymbol\kappa, \interior{\rm\bf K}_{ij}$
are freely specifiable throughout $\Sigma$. 

\subsection{The algebraic-hyperbolic constraint system}

An alternative approach is to
recast the Hamiltonian constraint as an algebraic equation for the scalar component
$\boldsymbol\kappa$ of $K_{ij}$.  The tangential derivatives of 
$\boldsymbol\kappa$ appearing in the momentum constraint for
${\rm\bf k}{}_{i}$ are then eliminated in terms of other variables.
This results in the momentum constraint system 
\begin{align} 
\mycal{L}_{\widehat n}({\rm\bf K}^l{}_{l}) - \widehat D^l {\rm\bf k}_{l} 
+ 2\,\dot{\widehat n}{}^l\, {\rm\bf k}_{l}
- [\,\boldsymbol\kappa-\tfrac12\, ({\rm\bf K}^l{}_{l})\,]\,
({\widehat K^{l}}{}_{l})  + \interior{\rm\bf K}{}_{kl}{\widehat K}{}^{kl}  = {}& 0 \,,   
\label{constr_mom2}  \\
\mycal{L}_{\widehat n} {\rm\bf k}{}_{i}  
+ ({\rm\bf K}^l{}_{l})^{-1}[\,\boldsymbol\kappa\,\widehat D_i ({\rm\bf K}^l{}_{l})
-2\, {\rm\bf k}{}^{l}\widehat D_i{\rm\bf k}{}_{l}\,] 
+ (2\,{\rm\bf K}^l{}_{l})^{-1}\widehat D_i\boldsymbol\kappa_0 {}& \nonumber \\ 
+ ({\widehat K^{l}}{}_{l})\,{\rm\bf k}{}_{i}
+ [\,\boldsymbol\kappa-\tfrac12\, ({\rm\bf K}^l{}_{l})\,]\,\dot{\widehat n}{}_i  
- \dot{\widehat n}{}^l\,\interior{\rm\bf K}_{li}  
+ \widehat D^l \interior{\rm\bf K}{}_{li}  = {}& 0 \, , \label{constr_mom1}
\end{align}
whereas the Hamiltonian constraint determines $\boldsymbol\kappa$
algebraically by 
\begin{equation} \label{constr_ham_n} 
\boldsymbol\kappa= (2\,{\rm\bf K}^l{}_{l})^{-1}[\, 2\,{\rm\bf k}{}^{l}{\rm\bf k}{}_{l} 
- \tfrac12\,({\rm\bf K}^l{}_{l})^2 - \boldsymbol\kappa_0 \,] \,,
\end{equation}
where
\begin{equation} \label{constr_ham_n0} 
\boldsymbol\kappa_0= {}^{{}^{(3)}}\hskip-1mm R 
- \interior{\rm\bf K}{}_{kl}\,\interior{\rm\bf K}{}^{kl}\,.
\end{equation}
Again, $\widehat D_i$ and $\widehat R$ denote the covariant derivative operator
and scalar curvature associated with
$\widehat \gamma_{ij}$, respectively, and $\dot{\widehat n}{}_k
={\widehat n}{}^lD_l{\widehat n}{}_k=-{\widehat D}_k(\ln{\widehat N})$.
(For more details see \cite{racz_constraints}).

By virtue of (\ref{constr_mom2})-(\ref{constr_ham_n0}) the four basic variables
$\boldsymbol\kappa, {\rm\bf k}{}_{i}, {\rm\bf K}^l{}_{l}$ are
determined by the constraints while the remaining eight
variables, represented by the 3-metric $h_{ij}$, consisting
of $(\widehat N,\widehat N^i,\widehat \gamma_{ij})$,
and $\interior{\rm\bf K}_{ij}$, are 
freely specifiable throughout $\Sigma$. As a result,
$\boldsymbol\kappa_0$ is determined by the freely
specified variables.
The constraint
system (\ref{constr_mom2})--(\ref{constr_ham_n0}) is
symmetric hyperbolic subject to the inequality
$\boldsymbol\kappa {\rm\bf K}^l{}_{l} <0$.

\section{Foliations by spheres or planes}
\label{sec:foliations}
\setcounter{equation}{0}

Two simple choices of foliations in solving  the
parabolic-hyperbolic system (\ref{bern_pde})--(\ref{ort_const_n})
or the algebraic-hyperbolic system (\ref{constr_mom2})--(\ref{constr_ham_n}) are
by spheres or planes, with tangential derivatives
referred to a background unit sphere geometry or
a background Euclidean geometry, respectively.

\subsection{Foliations by spheres and use of the $\eth$ operator}
\label{ethBAReth}

If we chose a foliation by two-spheres it is natural to solve the constraint equations
by decomposing the basic variables in terms of spin-weighted fields.
In doing so, the angular derivatives are expressed in terms of the
Newman-Penrose $\eth$ and $\,\overline\eth$
operators \cite{newman_penrose,goldberg_et_al}. We use
the notation and conventions introduced in \cite{jeff_edth,jeff_edth_2}
throughout this paper.

The metric $q_{ab}$ on the unit sphere $\mathbb{S}^2$, given in
standard $(\theta,\phi)$ coordinates by 
\begin{equation}\label{le}
ds^2= q_{ab}\, \d x^a \d x^b = \d\theta^2 + \sin^2\theta\, \d\phi^2\, ,
\end{equation}
provides a natural background geometry.
In terms of the complex stereographic coordinate
\begin{equation}
z = e^{-i\,\phi}\cot\frac{\theta}2=z_1+\mathbbm{i}\,z_2\,,
\end{equation}
the line element (\ref{le}) takes the form\,\footnote{Only expressions for the southern
hemisphere will  be given explicitly. Those on the northern hemisphere can be deduced
by the substitution $z_N=1/z_S$ \cite{jeff_edth,jeff_edth_2}.}
\begin{equation}\label{conf_flat}
ds^2 = 4\,(1+ z \,\overline z)^{-2}\left[\,(\d z_1)^2+(\d z_2)^2\,\right]\,.
\end{equation}

On $\mathbb{S}^2$, we choose the complex dyad
\begin{equation}\label{dyad}
q^a = {2^{-1}}{P}\left[\,(\partial_{z_1})^a + \mathbbm{i}\,(\partial_{z_2})^a\right]
= P\left(\partial_{\,\overline z}\right)^a\,,
\end{equation}
where 
\begin{equation}
P = 1+z\,\overline z \,.
\end{equation}
We also have
\begin{equation}\label{dual}
q_a = q_{ab}\,q^b = 2\, P^{-1}\left[\,(\d z_1)_a + \mathbbm{i}\,(\d z_2)_a\right]
= 2\,P^{-1}\left(\,\d z\right)_a\,.
\end{equation}

Note that $q^a$ has normalization
\begin{equation}\label{unit_metr-norm}
q^a \,\overline q_a =2\,, \quad q^a q_a =0\,, 
\end{equation}
and that the unit sphere metric $q_{ab}$ satisfies
\begin{equation}\label{unit_metr}
q _{ab} = q_{(a} \,\overline q_{b)}\,,  
\quad q^{ab} = q^{(a} \,\overline q{}^{\,b)}\,, \quad q^{ae} q_{eb} =\delta^a{}_b\,.
\end{equation}
In these conventions, the area element
on $\mathbb{S}^2$ is
$\boldsymbol\epsilon_{ab}=i\,q_{[a} \,\overline q_{b]}$. 

The Newman-Penrose $\eth$ and $\,\overline\eth$ operators are (see e.g.~(A4)
in \cite{jeff_edth})
\begin{align}
\eth\,\mathbb{L} = {}& P^{1-s}\,\partial_{\,\overline z} \left( P^s\,\mathbb{L} \right) 
\label{eth-def1} \\
\,\overline\eth\,\mathbb{L} = {}& P^{1+s}\,\partial_{z} \left( P^{-s}\,\mathbb{L} \right) \,,
\label{eth-def2}
\end{align}
where $\mathbb{L}$ is a spin-weight $s$ field
on $\mathbb{S}^2$ defined by 
\begin{equation}\label{eth-def3}
\mathbb{L}=q^{a_1}\dots q^{a_s}\, \mathbf{L}_{{a_1}\dots{a_s}} 
\end{equation}
for some totally symmetric traceless tensor field 
$\mathbf{L}_{{a_1}\dots{a_s}}$. 

As pointed out in \cite{jeff_edth,jeff_edth_2}, this choice of $\eth$ and
$\,\overline\eth$ corresponds to the
standard conventions in \cite{newman_penrose, goldberg_et_al,jeff_edth_2}.
The action of $\eth$ and
$\,\overline\eth$ on spin-weighted spherical harmonics ${}_{s}\mathbb{Y}_{\,l,m}$ is
given by  (see e.g.~(2.6)--(2.8) in \cite{goldberg_et_al})
\begin{align}
\eth\,{}_{s}\mathbb{Y}_{\,l,m}= {}& \sqrt{(l-s)(l+s+1)}\,{}_{s+1}\mathbb{Y}_{\,l,m} \, , \\
\,\overline\eth\,{}_{s}\mathbb{Y}_{\,l,m}= {}& -\sqrt{(l+s)(l-s+1)}\,{}_{s-1}\mathbb{Y}_{\,l,m} \, ,\\
\,\overline\eth\,\eth\,{}_{s}\mathbb{Y}_{\,l,m}= {}& -(l-s)(l+s+1)\,{}_{s}\mathbb{Y}_{\,l,m}\, ,
\end{align}
with
\begin{equation}
{}_{-s}\mathbb{Y}_{\,l,m}= {} (-1)^{s-m}\,\overline{{}_{s}\mathbb{Y}_{\,l,-m}} \, .
\end{equation}
\medskip

The $\eth$ and $\,\overline\eth$ operators are related to the covariant derivative
operator $\mathbb{D}_a$ associated with $q_{ab}$ by
\begin{align}
\eth\,\mathbb{L} = {}& q^b q^{a_1}\dots q^{a_s}\, \mathbb{D}_b\mathbf{L}_{({a_1}\dots{a_s})}\,,\\
\,\overline\eth\,\mathbb{L} = {}& {\,\overline q}^b q^{a_1}\dots q^{a_s}\, \mathbb{D}_b\mathbf{L}_{({a_1}\dots{a_s})}\,.
\end{align}
For a spin-weight $s$ field $\mathbbm{f}$, the commutation relation for covariant derivatives on $\mathbb{S}^2$ implies
$\left[\,\overline\eth, \eth\,\right]\,\mathbbm{f} = 2\,s\,\mathbbm{f}$. 

\subsection{Foliations by planes and the
related $\boldsymbol\partial$ and $\,\overline{\boldsymbol\partial}$ operators}
\label{papb}

In the formulation of a numerical algorithm based
upon a foliation by planes it is expedient to introduce
a background flat metric $q_{ab}$, analogous
to the previous treatment of spheres. Accordingly, the reference two-metric $q_{ab}$ on the
Euclidean plane $\mathbb{R}^2$ has decomposition
\begin{equation}\label{eq:flat_metric}
q _{ab} = q_{(a} \,\overline q_{b)}\,,  
\quad q^{ab} = q^{(a} \,\overline q{}^{\,b)}\,, \quad q^{ae} q_{eb} =\delta^a{}_b\,
\end{equation} 
in terms of the complex dyad 
\begin{equation}\label{eq:dual-dyad}
q_a = (\d x)_a + \mathbbm{i}\,(\d y)_a\,, \quad q^a
 = (\partial_{x})^a + \mathbbm{i}\,(\partial_{y})^a\,,
\end{equation}
with normalization
\begin{equation}\label{eq:flat_metric-norm}
q^a \,\overline q_a =2\,, \quad q^a q_a =0\,. 
\end{equation}

The field 
\begin{equation}\label{parc-def3}
\mathbb{L}=q^{a_1}\dots q^{a_s}\, \mathbf{L}_{{a_1}\dots{a_s}}\,,
\end{equation} 
where $\mathbf{L}_{{a_1}\dots{a_s}}$ is a symmetric traceless tensor field on $\mathbb{R}^2$,
has spin-weight $s$ with respect to rotations
of the dyad. Using standard complex notation,
analogs of the $\eth$ and $\overline{\eth}$ operators can be defined in terms of the flat covariant derivative operator associated with $q_{ab}$, i.e. the partial derivative $\partial_a$ with respect to Cartesian coordinates  $(x,y)$. This leads to the operators
\begin{align}
\boldsymbol\partial\,\mathbb{L} = {}& q^b q^{a_1}\dots q^{a_s}\, 
\partial_b\mathbf{L}_{({a_1}\dots{a_s})}\,,\\
\,\overline{\boldsymbol\partial}\,\mathbb{L} 
= {}& {\,\overline q}^b q^{a_1}\dots q^{a_s}\, \partial_b\mathbf{L}_{({a_1}\dots{a_s})}\,.
\end{align}

\subsection{The global  property of the spin-weighted
formalism}
\label{gdbf}

Consider one of the level surfaces $\mycal{S}_{\rho_0}$ of the foliation $\mycal{S}_\rho$.
If  $\mycal{S}_{\rho_0}$ is 
diffeomorphic to the sphere $\mathbb{S}^2$ we may introduce standard spherical
coordinates $(\theta,\phi)$ and the unit sphere metric (\ref{le}) on $\mycal{S}_{\rho_0}$.
Analogously, if  $\mycal{S}_{\rho_0}$ is 
diffeomorphic to $\mathbb{R}^2$, Cartesian
coordinates $(x,y)$ and a Euclidean metric
can be introduced.
In either case, by Lie dragging these coordinates
onto the leaves of the foliation $\mycal{S}_\rho$
by the evolution vector field $\rho^i$, their
values remain constant along the integral curves of $\rho^i$. By this process, either in terms of
the coordinates $(\theta,\phi)$ or $(x,y)$,
the corresponding metric $q_{ab}$ and complex dyad $q^a$ is defined on each
of the level surfaces $\mycal{S}_\rho$. 

\section{The spin-weight decomposition of the basic fields}\label{decomps}
\setcounter{equation}{0}

As a consequence of the above construction,
not only the coordinates $(\theta,\phi)$ or $(x,y)$ but  also  the complex dyad and
reference metric $q_{ab}$ are Lie dragged from $\mycal{S}_{\rho_0}$ onto the
surfaces  $\mycal{S}_\rho$, i.e. 
\begin{equation}\label{lie_dragged}
\mycal{L}_\rho\, q^a=0\,, \quad  \mycal{L}_\rho\, q_a=0
 \quad {\rm  and} \quad \mycal{L}_\rho\, q_{ab}=0  \,.
\end{equation}
It is a convenient consequence of this construction that
there is a single common treatment of the two cases based on foliations by
spheres or planes. Either case can be derived from the other by the simple
replacements $(\theta,\phi) \longleftrightarrow (x,y)$ and
$\eth \longleftrightarrow \boldsymbol{\partial}$. Accordingly, the
calculations below will be presented exclusively
for the case of a spherical foliation. 

\subsection{The decomposition of the metric $\widehat\gamma_{ab}$}

The metric $\widehat\gamma_{ab}$ induced on the $\mycal{S}_\rho$
level surfaces can be decomposed as
\begin{equation}\label{ind_metr}
\widehat\gamma_{ab}=\mathbbm{a}\, q_{ab}+\interior\gamma_{ab}\,, 
\end{equation}
where 
\begin{equation}
\mathbbm{a}=\tfrac12\,\widehat\gamma_{ab}\,q^a \,\overline q^b
\end{equation}
is a positive, spin-weight zero function on $\mycal{S}_\rho$ and
$\interior\gamma_{ab}$ is its trace-free part, i.e.
\begin{equation}\label{ind_metr_trf}
\interior\gamma_{ab}=\left[\delta_a{}^e\delta_b{}^f-\tfrac12\,q_{ab}\,q^{ef}\right]
  \widehat\gamma_{ef}=\widehat\gamma_{ab}-\mathbbm{a}\, q_{ab}\,.
\end{equation}

Since $\interior\gamma_{ab}$ is symmetric
and trace-free it has the decomposition
\begin{equation}\label{ind_metr_int}
\interior\gamma_{ab}=\tfrac12\left[ \mathbbm{b} \,\overline q_a \,\overline q_b 
+ \,\overline{\mathbbm{b}}\, q_a q_b \right] 
\end{equation}
in terms of the spin-weight $2$ function
\begin{equation}
\mathbbm{b}=\tfrac12\,\widehat\gamma_{ab}\,q^a q^b
=\tfrac12\,\interior\gamma_{ab}\,q^a q^b\,.
\end{equation}
The inverse metric has the decomposition
\begin{equation}\label{inv_ind_metr}
\widehat\gamma^{ab}=\mathbbm{d}^{-1}\left\{\mathbbm{a}\, q^{ab}
-\tfrac12\left[ \mathbbm{b} \,\overline q^a \,\overline q^b
+ \,\overline{\mathbbm{b}}\, q^a q^b \right]\right\}\,, 
\end{equation}
where 
\begin{equation}
\mathbbm{d}=\mathbbm{a}^2-\mathbbm{b}\,\overline{\mathbbm{b}}
\end{equation}
is the ratio  $\det(\widehat\gamma_{ab})/\det(q_{ab})$ of the determinants of
$\widehat\gamma_{ab}$ and $q_{ab}$. 

As an immediate application, using (\ref{unit_metr}), (\ref{ind_metr})
and (\ref{ind_metr_trf}), along with the notation
\begin{equation}
{\mathbbm{k}}=q^l\,{\rm\bf k}{}_{l}\,,\quad \,\overline{\mathbbm{k}}
=\,\overline q^l\,{\rm\bf k}{}_{l}\,,
\end{equation} 
${\rm\bf k}{}^{l} {\rm\bf k}{}_{l}$ can be expressed as 
\begin{align}
{\rm\bf k}{}^{l} {\rm\bf k}{}_{l}= {}& \widehat\gamma^{kl}{\rm\bf k}{}_{k}{\rm\bf k}{}_{l}
  =\tfrac12\, \mathbbm{d}^{-1}\left\{\mathbbm{a} \left( q^k\,\overline q^l 
  + q^l\,\overline q^k\right) - \left[\,\mathbbm{b}\, \,\overline q^k\,\overline q^l 
 + \,\overline{\mathbbm{b}}\, q^l q^k\, \right] \right\}{\rm\bf k}{}_{k}{\rm\bf k}{}_{l} 
    \nonumber \\ 
 = {}& \tfrac12\,\,\mathbbm{d}^{-1} [\, 2\,\mathbbm{a}\,{\mathbbm{k}}\,\overline{\mathbbm{k}}
 - \mathbbm{b}\,\overline{\mathbbm{k}}^2- \,\overline{\mathbbm{b}}\,\mathbbm{k}^2\,]\,.
\end{align}

\subsection{Terms involving the covariant derivative $\widehat D_{a}$}

The covariant derivative operators $\widehat D_{a}$ and ${\mathbb D}_{a}$ can be related
by the tensor field
\begin{align}
{C^e}{}_{ab}=  \tfrac12\,\widehat\gamma^{ef}\left\{ {\mathbb D}_{a}\widehat\gamma_{fb}
+ {\mathbb D}_{b}\widehat\gamma_{af}-{\mathbb D}_{f}\widehat\gamma_{ab}\right\} \,. 
\end{align}
(See e.g.~(3.1.28) and (D.3) in \cite{wald}.) 
In particular,
\begin{equation}
\widehat D_{a} {\rm\bf k}{}_{b}={\mathbb D}_{a} {\rm\bf k}{}_{b}
 -{C^e}{}_{ab}  {\rm\bf k}{}_{e} \,  ,
\end{equation}
and thereby 
\begin{align}\label{divk}
             \widehat D^{l} \,{\rm\bf k}{}_{l} 
             = {}& \widehat\gamma^{kl} \,\widehat D_{k} \,{\rm\bf k}{}_{l}
             =\tfrac12\,\mathbbm{d}^{-1}\left\{\mathbbm{a} \left( q^k\,\overline q^l
              + q^l\,\overline q^k\right) 
             - \left[\,\mathbbm{b}\, \,\overline q^k\,\overline q^l
              + \,\overline{\mathbbm{b}}\, q^l q^k\, \right] \right\} \widehat D_{k} \,{\rm\bf k}{}_{l} 
              \nonumber \\
            = {}& \tfrac14\,\mathbbm{d}^{-1}\left\{ 2 \mathbbm{a}
             \left(\,\eth\,\overline{\mathbbm{k}}-\mathbb{B}\,\overline{\mathbbm{k}}  \right) 
             - \mathbbm{b} \left(2\,\overline{\eth}\,\overline{\mathbbm{k}}
             -\,\overline{\mathbb{C}}\,\mathbbm{k}
             -\,\overline{\mathbb{A}}\,\overline{\mathbbm{k}}  \right) 
             + ``\,CC\,"\right\} \,,
\end{align}
where 
\begin{align} \label{ABC}
\mathbb{A} = {}&  q^a q^b {C^e}{}_{ab}\,\overline q_e 
= \mathbbm{d}^{-1}\left\{ \mathbbm{a}\left[2\,\eth\,\mathbbm{a} 
-\,\overline{\eth}\,\mathbbm{b}\right] 
   -  \,\overline{\mathbbm{b}}\,\eth\,\mathbbm{b} \right\} \nonumber \\
\mathbb{B} = {}&  \,\overline q^a q^b {C^e}{}_{ab}\,q_e 
  = \mathbbm{d}^{-1}\left\{ \mathbbm{a}\,\overline{\eth}\,\mathbbm{b}
  - \mathbbm{b}  \,\eth\,\overline{\mathbbm{b}}\right\} \\
\mathbb{C} = {}&  q^a q^b {C^e}{}_{ab}\,q_e 
  = \mathbbm{d}^{-1}\left\{ \mathbbm{a}\,\eth\,\mathbbm{b} 
  -  \mathbbm{b}\left[2\,\eth\,\mathbbm{a}
   -\,\overline{\eth}\,\mathbbm{b}\right] \right\} \,. \nonumber
\end{align}
Hereafter $``\,CC\,"$ stands for the complex conjugate of the previous terms
within the same brackets or parentheses. 

We also obtain the relation 
\begin{align}\label{2kDk}
\hskip-1cm
     [\,2\,{\rm\bf k}{}^{l}\widehat D_{i} \,{\rm\bf k}{}_{l}\,]\,q^i
     = & [\,2\,\widehat\gamma^{kl}{\rm\bf k}{}_{k}\,{\widehat D}_{i} \,{\rm\bf k}{}_{l} \,]\,q^i  \\ 
     = {}& [\,\mathbbm{d}^{-1}\left\{\mathbbm{a} \left( q^k\,\overline q^l 
     + q^l\,\overline q^k\right) - \left[\,\mathbbm{b}\, \,\overline q^k\,\overline q^l 
     + \,\overline{\mathbbm{b}}\, q^l q^k\, \right] \right\} {\rm\bf k}{}_{k}
     \,{\widehat D}_{i} \,{\rm\bf k}{}_{l}\, ]\,q^i \nonumber\\ 
     =  \tfrac12\,\,\mathbbm{d}^{-1}\left\{ \left(\mathbbm{a} \,\mathbbm{k}
     -\mathbbm{b}\,\overline{\mathbbm{k}}\right)\right.{}
     &\hskip-2mm\left. \left[2\,\eth\,\overline{\mathbbm{k}}-\,\overline{\mathbb{B}}\,
     \mathbbm{k}-\mathbb{B}\,\overline{\mathbbm{k}}  \right] 
     + \left(\mathbbm{a} \,\overline{\mathbbm{k}}
     -\,\overline{\mathbbm{b}}\,\mathbbm{k}\right)\left[2\,\eth\,\mathbbm{k}
     -\mathbb{C}\,\overline{\mathbbm{k}}-\mathbb{A}\,\mathbbm{k}  \right] \right\} .
      \nonumber
\end{align}

\subsection{The scalar curvature ${}^{{}^{(3)}}\hskip-1mm R$}

In expressing the scalar curvature   ${}^{{}^{(3)}}\hskip-1mm R$ in terms
of spin-weighted fields we use the relation
\begin{equation}\label{R3}
{}^{{}^{(3)}}\hskip-1mm R= \widehat R - [\,2\,\mycal{L}_{\widehat n} ({\widehat K^l}{}_{l}) 
   + ({\widehat K^{l}}{}_{l})^2 + \widehat K_{kl} \widehat K^{kl}
  + 2\,{\widehat N}^{-1}\,\widehat D^l \widehat D_l \widehat N \,]\,,
\end{equation}
where $\widehat R$ is the scalar curvature of the
metric $\widehat\gamma_{ab}$, given by
\begin{equation}\label{R2}
\widehat R = \,\widehat{\mathbb{R}} = \tfrac12\, {\mathbbm{a}}^{-1}\left( \mathbb{R}
- \left\{ \,  \eth\,\overline{\mathbb{B}} - \overline{\eth}\,\mathbb{A}
- \tfrac12\,\left[\, \mathbb{C}\,\overline{\mathbb{C}}  
- \mathbb{B}\,\overline{\mathbb{B}} \,\right]\, \right\}\,\right) \,
\end{equation}
in terms of the scalar curvature $\mathbb{R}$ associated with ${\mathbb D}_{a}$.
(For the foliation by spheres $\mathbb{R}=2$, and for planes $\mathbb{R}=0$.)
The basic field variables used in recasting the constraint equations are collected
in Table\,\ref{table:data}.  
\begin{table}[]
	\centering  \hskip-.15cm
	\begin{tabular}{|c|c|c|} 
		\hline notation &  definition  & \hskip-0.7cm$\phantom{\frac{\frac12}{A}_{B_D}}$ 
		spin-weight \\ \hline \hline
		
		$\mathbbm{a}$ &  $\tfrac12\,q^i\,\overline q^j\,\widehat\gamma_{ij}$  & \hskip-0.7cm$  
		\phantom{\frac{\frac12}{A}_{B_D}}$ $0$ \\  \hline 
		
		$\mathbbm{b}$ &  $\tfrac12\,q^i q^j\,\widehat\gamma_{ij}$ 
		 & \hskip-0.7cm$\phantom{\frac{\frac12}{A}_{B_D}}$ $2$ \\  \hline 
		
		$\mathbbm{d}$ &  $\mathbbm{a}^2-\mathbbm{b}\,\overline{\mathbbm{b}}$  
		& \hskip-0.7cm$\phantom{\frac{\frac12}{A}_{B_D}}$ $0$ \\  \hline 
		$\mathbbm{k}$ &  $q^i {\rm\bf k}{}_{i}$  & 
		\hskip-0.7cm$\phantom{\frac{\frac12}{A}_{B_D}}$ $1$
		 \\  \hline 
		
		$\mathbb{A}$ &  $q^a q^b {C^e}{}_{ab}\,\overline q_e
		= \mathbbm{d}^{-1}\left\{ \mathbbm{a}\left[2\,\eth\,\mathbbm{a}
		-\,\overline{\eth}\,\mathbbm{b}\right] 
		-  \,\overline{\mathbbm{b}}\,\eth\,\mathbbm{b} \right\} $ 
		 & \hskip-0.7cm$\phantom{\frac{\frac12}{A}_{B_D}}$ $1$ \\  \hline 
		
		$\mathbb{B}$ &  $\,\overline q^a q^b {C^e}{}_{ab}\,q_e
		 = \mathbbm{d}^{-1}\left\{ \mathbbm{a}\,\overline{\eth}\,\mathbbm{b}
		- \mathbbm{b}  \,\eth\,\overline{\mathbbm{b}}\right\}$ 
		& \hskip-0.7cm$\phantom{\frac{\frac12}{A}_{B_D}}$ $1$ \\  \hline 
		
		$\mathbb{C}$ &  $q^a q^b {C^e}{}_{ab}\,q_e 
		= \mathbbm{d}^{-1}\left\{ \mathbbm{a}\,\eth\,\mathbbm{b}
		 -  \mathbbm{b}\left[2\,\eth\,\mathbbm{a} 
		-\,\overline{\eth}\,\mathbbm{b}\right] \right\}$ 
		& \hskip-0.7cm$\phantom{\frac{\frac12}{A}_{B_D}}$ $3$ \\  \hline 
		
		$\,\widehat{\mathbb{R}}$ &  $\tfrac12\, {\mathbbm{a}}^{-1}\left( \mathbb{R}
		- \left\{ \,  \eth\,\overline{\mathbb{B}} - \overline{\eth}\,\mathbb{A}
		- \tfrac12\,\left[\, \mathbb{C}\,\overline{\mathbb{C}} 
		 - \mathbb{B}\,\overline{\mathbb{B}} \,\right]\, \right\}\,\right)$ 
		 & \hskip-0.7cm$\phantom{\frac{\frac12}{A}_{B_D}}$ $0$ \\  \hline 
		
		$\,\widehat{\mathbb{N}}$ &  $\widehat N$  
		& \hskip-0.7cm$\phantom{\frac{\frac12}{A}_{B_D}}$ $0$ \\  \hline
		$\mathbb{N}$ &  $q^i\widehat N_i= q^i \widehat\gamma{}_{ij} {\widehat N}{}^{j}$ 
		 & \hskip-0.7cm$\phantom{\frac{\frac12}{A}_{B_D}}$ $1$ \\  \hline
		
		$\widetilde{\mathbb{N}}$ &  $q_i\widehat N^i 
		= q_i \,\widehat\gamma{}^{ij} {\widehat N}{}_{j}
		=\mathbbm{d}^{-1} (\mathbbm{a}\,\mathbb{N} - \mathbbm{b}\,\overline{\mathbb{N}})$  
		& \hskip-0.7cm$\phantom{\frac{\frac12}{A}_{B_D}}$ $1$ \\  \hline
		
		$\mathbb{K}$ &  $ \widehat\gamma^{kl} \,{\rm\bf K}{}_{kl}
		$  & \hskip-0.7cm$\phantom{\frac{\frac12}{A}_{B_D}}$ $0$ \\  \hline 
		
		$\interior{\mathbb{K}}{}_{qq}$ &  $q^kq^l\,\interior{\rm\bf K}{}_{kl}
		$  & \hskip-0.7cm$\phantom{\frac{\frac12}{A}_{B_D}}$ $2$ \\  \hline 
		
		$\interior{\mathbb{K}}{}_{q\overline{q}}$ &  $q^k\,\overline{q}^l\,\interior{\rm\bf K}{}_{kl}
		= (2\,\mathbbm{a})^{-1} [\,\mathbbm{b}\,\overline{\interior{\mathbb{K}}{}_{qq}} 
		+  \overline{\mathbbm{b}}\,\interior{\mathbb{K}}{}_{qq} \,] 
		$  & \hskip-0.7cm$\phantom{\frac{\frac12}{A}_{B_D}}$ $0$ \\  \hline   
		
		$\,\widehat{\mathbb{K}}$ &  ${\widehat K}^l{}_{l} = \widehat\gamma^{ij} \widehat K_{ij} $ 
		 & \hskip-0.7cm$\phantom{\frac{\frac12}{A}_{B_D}}$ $0$ \\  \hline 
		
		$\widehat{\mathbb{K}}{}_{qq}$ &  $q^i q^j\widehat K_{ij} 
		= \tfrac12\,{\,\widehat{\mathbb{N}}}^{-1}\left\{2\,\partial_\rho\mathbbm{b} - 2\,\eth\,\mathbb{N}
		+ {\mathbb{C}}\,\overline {\mathbb{N}} +\mathbb{A} \,{\mathbb{N}} \,  \right\} $ 
		& \hskip-0.7cm$\phantom{\frac{\frac12}{A}_{B_D}}$ $2$ \\  \hline

		$\widehat{\mathbb{K}}{}_{q\overline{q}}$ &  $q^k\,\overline{q}^l\,\widehat{K} {}_{kl} 
		= {\mathbbm{a}}^{-1}\{\,\mathbbm{d}\cdot\widehat{\mathbb{K}} 
		+ \tfrac12 \,[\,\mathbbm{b}\,\overline{\widehat{\mathbb{K}}{}_{qq}} 
		+  \overline{\mathbbm{b}}\,\widehat{\mathbb{K}}{}_{qq}  \,]\,\}
		$  & \hskip-0.7cm$\phantom{\frac{\frac12}{A}_{B_D}}$ $0$ \\  \hline  
		
		$\,\instar{\mathbb{K}}$ &  ${\instar{K}}{}^l{}_{l} = \widehat\gamma^{ij} \instar{K}_{ij} $  
		& \hskip-0.7cm$\phantom{\frac{\frac12}{A}_{B_D}}$ $0$ \\  \hline 
		
		$\instar{\mathbb{K}}{}_{qq}$ &  $q^i q^j\instar K_{ij} 
		= \tfrac12\,\left\{2\,\partial_\rho\mathbbm{b} - 2\,\eth\,\mathbb{N}
		+ {\mathbb{C}}\,\overline {\mathbb{N}} +\mathbb{A} \,{\mathbb{N}} \,  \right\} $ 
		& \hskip-0.7cm$\phantom{\frac{\frac12}{A}_{B_D}}$ $2$ \\  \hline   
		
		$\instar{\mathbb{K}}{}_{q\overline{q}}$ &  $q^k\,\overline{q}^l\,\instar{K} {}_{kl} 
		=  {\mathbbm{a}}^{-1}\{\,\mathbbm{d}\cdot\instar{\mathbb{K}} 
		+ \tfrac12 \,[\,\mathbbm{b}\,\overline{\instar{\mathbb{K}}{}_{qq}} 
		+  \overline{\mathbbm{b}}\,\instar{\mathbb{K}}{}_{qq}  \,]\,\}
		$  & \hskip-0.7cm$\phantom{\frac{\frac12}{A}_{B_D}}$ $0$ \\  \hline                  
		
	\end{tabular}
	\caption{\small The new variables used in recasting the constraints.
	For detailed derivations of these and other complicated expressions
	see the Appendix. }
	\label{table:data}
\end{table}

\section{The constraints in terms of the new variables}
\label{sec:newvar}
\setcounter{equation}{0}

We now present the explicit form of the constraints in terms of the spin-weighted fields
and their derivatives introduced in the previous sections.

\subsection{The parabolic-hyperbolic system}

Application of the spin-weight decomposition of the
basic variables leads to the following form
of the parabolic-hyperbolic system (\ref{bern_pde})--(\ref{ort_const_n}), 
\begin{align} 
{}& \instar{\mathbb{K}}\,[\,\partial_{\rho} \widehat{\mathbb{N}}
 - \tfrac12\,\widetilde{\mathbb{N}} \,(\,\overline{\eth}\, \widehat{\mathbb{N}}) 
 -\tfrac12\, \,\overline{\widetilde{\mathbb{N}}}
\, (\eth\,\widehat{\mathbb{N}}) \,] \label{bern_pde2}\\ {}& 
- \tfrac12\,{\mathbbm{d}}^{-1}\widehat{\mathbb{N}}^{\,2}[\,\mathbbm{a}
 \{\,(\eth\,\overline{\eth}\,\widehat{\mathbb{N}}) 
- \mathbb{B}\,(\,\overline{\eth}\,\widehat{\mathbb{N}}) \,\}  
 - \mathbbm{b} \,\{\,(\,\overline{\eth}^2\,\widehat{\mathbb{N}}) 
- \tfrac12\,\overline{\mathbb{A}}\,(\,\overline{\eth}\,\widehat{\mathbb{N}}) 
- \tfrac12\,\overline{\mathbb{C}} \, ({\eth}\,\widehat{\mathbb{N}}) \,\} 
 + ``\,CC\,"  \,]\nonumber \\ {}& \hskip 10cm - \mathcal{A}\,\widehat{\mathbb{N}} - \mathcal{B}\,\widehat{\mathbb{N}}{}^{\,3}   =0 \,, \nonumber  \\ 
 {}&   \partial_\rho  {\mathbbm{k}}  
 - \tfrac12\,\widetilde{\mathbb{N}} \,(\,\overline{\eth}\, \mathbbm{k}) 
 -\tfrac12\, \,\overline{\widetilde{\mathbb{N}}}
 \, (\eth\,\mathbbm{k}) - \tfrac12 \,\widehat{\mathbb{N}}\,\eth\,\mathbb{K} 
 + \mathbbm{f}_{\mathbbm{k}} = 0 \label{eq:eth_constr_mom1}\,, \\
{}&  \partial_\rho \mathbb{K} - \tfrac12\,\widetilde{\mathbb{N}} \,(\,\overline{\eth}\, \mathbb{K}) 
 -\tfrac12\, \,\overline{\widetilde{\mathbb{N}}}
\, (\eth\,\mathbb{K}) - \tfrac12\,\widehat{\mathbb{N}}\,\mathbbm{d}^{-1}
 \left\{\, \mathbbm{a}\,(\eth\,\overline{\mathbbm{k}} + \,\overline{\eth}{\mathbbm{k}})
- \mathbbm{b}\,\overline{\eth}\,\overline{\mathbbm{k}} 
- \,\overline{\mathbbm{b}}\,\eth{\mathbbm{k}}\,\right\} + \mathbb{F}_{\mathbb{K}}
=0 \label{eq:eth_constr_mom2} \,.
\end{align}

In (\ref{eq:eth_constr_mom1})--(\ref{eq:eth_constr_mom2}), the lower order source terms
$\mathbbm{f}_{\mathbbm{k}}$ and  $\mathbb{F}_{\widehat{\mathbb{K}}}$ have
spin-weight $1$ and $0$, respectively, and the coefficients $\mathcal{A}$ and ${\mathcal{B}}$
have spin-weight $0$, on each surface $\mycal{S}_\rho$. They are smooth functions of
the constrained variables $\widehat{\mathbb{N}}, \mathbb{K}, \mathbbm{k}$ and the freely
specified variables
$\mathbbm{a},\mathbbm{b}, \mathbb{N}, \,\boldsymbol{\kappa}, \interior{\mathbb{K}}{}_{qq}$,
along with the $\eth$, $\overline{\eth}$ and $\rho$-derivatives of the
free variables. Their explicit forms are 
\begin{align}
\mathbbm{f}_{\mathbbm{k}} = {}
 & -\tfrac12 \left[\,  \mathbbm{k} \,\eth \,\overline{\widetilde{\mathbb{N}}}
+\overline {\mathbbm{k}}\, \eth \,{\widetilde{\mathbb{N}}} \right]  
- [\,\boldsymbol\kappa-\tfrac12\, \mathbb{K}\,]\,  \eth\,\widehat{\mathbb{N}} 
+ \mathbbm{p}  \nonumber \\ {}& 
+ \widehat{\mathbb{N}} \left[-\eth \boldsymbol\kappa
+ \widehat{\mathbb{N}}^{-1}\instar{\mathbb{K}} \,\mathbbm{k}
 -q^i \dot{\widehat n}{}^l\,\interior{\rm\bf K}_{li}  
+ q^i\widehat D^l \interior{\rm\bf K}{}_{li}  \right] \label{eq:bff} \, ,\\	
\mathbb{F}_{\mathbb{K}} = {} &	\tfrac14\,\widehat{\mathbb{N}}\,\mathbbm{d}^{-1}\left\{    
2 \, \mathbbm{a} \,\mathbbm{B} \,\overline{\mathbbm{k} }  
-  \mathbbm{b}\, (\, \overline{ \mathbbm{C}} \,\mathbbm{k}
+\overline{ \mathbbm{A}} \, \overline{\mathbbm{k}} \,)
- \left\{\mathbbm{p}_\rho - \tfrac12\,[\widetilde{\mathbb{N}}\,\overline{\mathbbm{p}} 
+ \overline{\widetilde{\mathbb{N}}}
\,\mathbbm{p}]\right\} 
+ ``\,CC\,"  \right \} \nonumber \\
&  \hskip-0.2cm -\mathbbm{d}^{-1} \left[ (\,\mathbbm{a} \,\overline  {\mathbbm{k}} 
-\overline {\mathbbm{b}}\, \mathbbm{k}\,)\,\eth\,\widehat{\mathbb{N}} 
+ ``\,CC\," \right]
+ \left[\,
\interior{\rm\bf K}{}_{ij}   {\instar{K}}{}^{ij} - (\,\boldsymbol{\kappa}
 -\tfrac12 \,\mathbb{K}\,)\,\instar{\mathbb{K}}\,\right]     \, ,   
\end{align}
\begin{align}
{}& \mathcal{A} =\partial_{\rho} \instar{\mathbb{K}} 
- \tfrac12\,\widetilde{\mathbb{N}} \,(\,\overline{\eth}\, \instar{\mathbb{K}}) -\tfrac12\, \,\overline{\widetilde{\mathbb{N}}}
\, (\eth\,\instar{\mathbb{K}}) + \tfrac{1}{2}[\,\instar{\mathbb{K}}^2
 + \instar{\mathbb{K}}{}_{kl} \instar{\mathbb{K}}{}^{kl}\,] \, , \label{A2}  \\
{}& \mathcal{B} =- \tfrac12\,\bigl[\,\widehat{\mathbb{R}} 
+ 2\,\boldsymbol\kappa\,\mathbb{K} + \tfrac12\,\mathbb{K}^2 
- \mathbbm{d}^{-1} [\, 2\,\mathbbm{a}\,{\mathbbm{k}}\,\overline{\mathbbm{k}}
- \mathbbm{b}\,\overline{\mathbbm{k}}^2- \,\overline{\mathbbm{b}}\,\mathbbm{k}^2\,] 
- \interior{\rm\bf K}{}_{kl}\,\interior{\rm\bf K}{}^{kl}\,\bigr] \label{eq:B2}\, ,
\end{align}
where the explicit forms of the new terms
introduced in \eqref{eq:bff}--\eqref{eq:B2} are
\begin{align}
q^{i\,} \dot{\widehat n}{}^{k\,} \interior{\rm\bf K}{}_{ki}	 = {} &  
-\tfrac12 ( {\widehat{\mathbb N}}\,{\mathbbm d})^{-1} \left[
\mathbbm a\,(\overline\eth\, \widehat{\mathbb N} ) \, \interior{\mathbb{K}}{}_{qq}    
+  \mathbbm a\,( \eth {\widehat{\mathbb N}} ) \, \interior{\mathbb{K}}{}_{q\overline{q}}  
-\mathbbm b\,(\overline\eth {\widehat{\mathbb N}} ) \, \interior{\mathbb{K}}{}_{q\overline{q}} 
- \overline {\mathbbm b}\,(\eth {\widehat{\mathbb N}} ) \,
\interior{\mathbb{K}}{}_{qq}   \right]\,, \label{5.8}\\
q^i \widehat D^{k\,} \interior{\rm\bf K}{}_{ki}	 = {} &
\frac{ 1} {2\mathbbm d}\left(
\mathbbm a\,\overline\eth \, \interior{\mathbb{K}}{}_{qq}    
+  \mathbbm a\,\eth \, \interior{\mathbb{K}}{}_{q\overline{q}}  
-\mathbbm b \,\overline\eth \, \interior{\mathbb{K}}{}_{q\overline{q}} 
- \overline {\mathbbm b}\, \eth \, \interior{\mathbb{K}}{}_{qq}   \right) 
+ \frac{\overline{ \mathbbm b}} {2\mathbbm d} \left( \mathbbm A\,\interior{\mathbb{K}}{}_{qq}  
+\mathbbm C\, \interior{\mathbb{K}}{}_{q\overline{q}}  \right)  \nonumber
\\
& - \frac{ \mathbbm a} {4\mathbbm d} \left(
3\,\overline{\mathbbm B} \,\interior{\mathbb{K}}{}_{qq} 
+3\,{\mathbbm B}\, \interior{\mathbb{K}}{}_{q\overline{q}} 
+{\mathbbm A} \,\interior{\mathbb{K}}{}_{q\overline{q}} +{\mathbbm C}\,
\overline{\interior{\mathbb{K}}{}_{qq}} \right)
\nonumber \\
& + \frac{ \mathbbm b} {4\mathbbm d} \left(
\overline{\mathbbm C} \,\interior{\mathbb{K}}{}_{qq} 
+\overline{\mathbbm A} \,\interior{\mathbb{K}}{}_{q\overline{q}} 
+\overline{\mathbbm B} \,\interior{\mathbb{K}}{}_{q\overline{q}}
 +{\mathbbm B} \,\overline{\interior{\mathbb{K}}{}_{qq}} \right) 
\,, \label{eq:divKnull} \\
\interior{\rm\bf K}{}_{ij} \instar{K}{}^{ij}  = {} & \tfrac14\, {\mathbbm d}^{-2} 
\left\{ \,2\, \interior{\mathbb{K}}{}_{q\overline{q}}\left[ \left(\,{\mathbbm a}^2
+{\mathbbm b}\,\overline {\mathbbm b}\right)\,{\instar{\mathbb{K}}{}_{q\overline{q}} } 
- {\mathbbm a}\left(\,{\overline{\mathbbm b}}\,{\instar{\mathbb{K}}{}_{qq}}  
+ {\mathbbm b} {\,\overline{\instar{\mathbb{K}}{}_{qq}} } \right)\,\right] \right.  \nonumber \\ 
{}& \left.
\hskip1.2cm
+ \left[ \,{\overline{\interior{\mathbb{K}}{}_{qq}}} 
\left(\, {\mathbbm a}^2\,{\instar{\mathbb{K}}{}_{qq}}  
+ {\mathbbm b}^2 {\,\overline{\instar{\mathbb{K}}{}_{qq}} } 
-2\,\mathbbm a \,\mathbbm b \,\instar{\mathbb{K}}{}_{q\overline{q}} \right)
+ ``\,CC\," \right] \right\}\,, \label{5.10} \\
\instar{K}{}_{ij} \instar{K}{}^{ij}  {}&  = \tfrac14\, {\mathbbm d}^{-2} \left\{
\left[ \,\overline{\,\instar{\mathbb{K}}{}_{qq}}
\left( {\mathbbm a}^2\,\instar{\mathbb{K}}{}_{qq} 
+ {\mathbbm b}^2 \,\overline{\,\instar{\mathbb{K}}{}_{qq} } 
-4\,\mathbbm a \,\mathbbm b \,\instar{\mathbb{K}}{}_{q\overline{q}} \right)
+ ``\,CC\," \right] \right. \nonumber  \\ {} & \left.  
\hskip6.8cm
+  2\, ({\mathbbm a}^2
+{\mathbbm b}\,\overline {\mathbbm b} )\,\instar{\mathbb{K}}{}_{q\overline{q}}^2\right\} \,, 
\label{5.11} \\
\interior{\rm\bf K}{}_{ij}   \interior{\rm\bf K}{}^{ij}  
{} & = \tfrac14  \,{\mathbbm d}^{-2} \left\{
\left[ \, {\overline { \interior{\mathbb{K}}{}_{qq}}} \,
(\, {\mathbbm a}^2\,{ \interior{\mathbb{K}}{}_{qq}}   
+ {\mathbbm b}^2  {\,\overline { \interior{\mathbb{K}}{}_{qq}}}  
-4\,\mathbbm a \,\mathbbm b \,\interior{\mathbb{K}}{}_{q\overline{q}}\, )
+ ``\,CC\," \right] \right. \nonumber \\ & \left.  
\hskip6.8cm
+  2\,  ({\mathbbm a}^2+{\mathbbm b}\,\overline {\mathbbm b} )
\,\interior{\mathbb{K}}{}_{q\overline{q}}^2\right\} \,.
\label{eq:abb-last}
\end{align}

For detailed derivation of these relations see the
Appendix.

\subsection{The algebraic-hyperbolic system}

Application of the spin-weight decomposition of the
basic variables introduced in Table\,\ref{table:data} leads
to the following form of the algebraic-hyperbolic constraints, 
\begin{align} 
          \partial_\rho \mathbb{K} {}& 
          - \tfrac12\,\widetilde{\mathbb{N}} \,(\,\overline{\eth}\, \mathbb{K})
           -\tfrac12\, \,\overline{\widetilde{\mathbb{N}}}
            \, (\eth\,\mathbb{K}) - \tfrac12\,\widehat{\mathbb{N}}\,\mathbbm{d}^{-1}
            \left\{\, \mathbbm{a}\,(\eth\,\overline{\mathbbm{k}} + \,\overline\eth{\mathbbm{k}})
             - \mathbbm{b}\,\overline\eth\,\overline{\mathbbm{k}} 
             - \,\overline{\mathbbm{b}}\,\eth{\mathbbm{k}}\,\right\}  \nonumber \\
             {}&+ \mathbb{F}_{\mathbb{K}}
             =  0 \, , \label{eq:eth2_constr_mom2} \\
     \partial_\rho  {\mathbbm{k}}    {}&  - \tfrac12\,\widetilde{\mathbb{N}} \,(\,\overline{\eth}\, \mathbbm{k}) 
     -\tfrac12\, \,\overline{\widetilde{\mathbb{N}}}
         \, (\eth\,\mathbbm{k}) + \,\widehat{\mathbb{N}}\,(\mathbb{K})^{-1}
         \left\{\,\boldsymbol{\kappa}\,(\eth\,\mathbb{K}) 
         - \,\,\mathbbm{d}^{-1}\,[\,(\mathbbm{a}\,\mathbbm{k}
         -\mathbbm{b}\,\overline{\mathbbm{k}})\,(\eth{\,\overline{\mathbbm{k}}}) \right.{}
          \label{eq:eth2_constr_mom1} \nonumber \\ 
            {}& \left. + (\mathbbm{a}\,\overline{\mathbbm{k}} -\,\overline{\mathbbm{b}}\,\mathbbm{k} )
            \,(\eth{\mathbbm{k}}) \,]  \right\} 
         + \mathbbm{f}_{\mathbbm{k}} =  0 \,,  \\
           {}& \boldsymbol\kappa= (2\,\mathbb{K})^{-1}\left[\,
               \mathbbm{d}^{-1}(  2\,\mathbbm{a}\,{\mathbbm{k}}\,\overline{\mathbbm{k}}
                - \mathbbm{b}\,\overline{\mathbbm{k}}^2
              - \,\overline{\mathbbm{b}}\,\mathbbm{k}^2)  - \tfrac12\,\mathbb{K}^2 
              - \boldsymbol\kappa_0 \,\right]\,,
              \label{eq:eth2_constr_ham_n} 
\end{align}
where, in virtue of (\ref{constr_ham_n0}), $\boldsymbol\kappa_0$ can be evaluated by
applying (\ref{R3}), (\ref{R2}), (\ref{tracefreebfcurv_sq}), (\ref{hatextcurv_trace}),
(\ref{hatextcurv_sq}) and (\ref{Lie_hatK}).

In (\ref{eq:eth2_constr_mom2})--(\ref{eq:eth2_constr_mom1}), the lower order source terms
$\mathbb{F}_{\mathbb{K}}$
and $\mathbbm{f}_{\mathbbm{k}}$ have spin-weight $0$ and $1$, respectively.
They are both smooth undifferentiated functions of the constrained variables
$\boldsymbol{\kappa},\mathbb{K}, \mathbbm{k}$; and they are also smooth functions of
the freely specifiable variables
$\mathbbm{a},\mathbbm{b},  \,\widehat{\mathbb{N}}, \mathbb{N},
 \interior{\mathbb{K}}{}_{qq}$
and their derivatives. The explicit forms of the forcing terms are
\begin{align}
            \mathbb{F}_{\mathbb{K}} = {} &
            	\tfrac14\,\widehat{\mathbb{N}}\,\mathbbm{d}^{-1}\left\{    
	         2 \, \mathbbm{a} \,\mathbbm{B} \,\overline{\mathbbm{k} }  
		 -  \mathbbm{b}\, (\, \overline{ \mathbbm{C}} \,\mathbbm{k}
		 +\overline{ \mathbbm{A}} \, \overline{\mathbbm{k}} \,)
		+ ``\,CC\,"  \right \} \\		
		&  -\mathbbm{d}^{-1} \left[ (\,
		\mathbbm{a} \,\overline  {\mathbbm{k}}
		-\overline {\mathbbm{b}}\, \mathbbm{k}\,)\,\eth\,\widehat{\mathbb{N}} 
		+ ``\,CC\," \right]
	 +\widehat{\mathbb{N}} \left[\,\interior{\rm\bf K}{}_{ij}   {\widehat K}{}^{ij}
	 -(\,\boldsymbol{\kappa} -\tfrac12 \,\mathbb{K}\,)
	 \,\widehat{\mathbb{K}}\,\right]  \, ,\nonumber \\ 
            \mathbbm{f}_{\mathbbm{k}} 
            = {}& -\tfrac12 \left[\,  \mathbbm{k} \,\eth \,\overline{\widetilde{\mathbb{N}}}
        +\overline {\mathbbm{k}}\, \eth \,{\widetilde{\mathbb{N}}} \right] \\ {}& 
            +\tfrac12 \,\widehat{\mathbb{N}}\,(\mathbbm{d}\,\mathbb{K})^{-1} 
            \left [\, (\mathbbm{a}\,\mathbbm{k}
            -\mathbbm{b}\,\overline{\mathbbm{k}})
            \,(\overline{\mathbb{B}}\,\mathbbm{k}
        +\mathbb{B}\,\overline{\mathbbm{k}}) 
         + (\mathbbm{a}\,\overline{\mathbbm{k}} 
         -\,\overline{\mathbbm{b}}\,\mathbbm{k} )
         \,(\mathbb{C}\, \overline{\mathbbm{k}}
         +\mathbbm{A}\,\mathbbm{k})\, \right ] \, , \nonumber \\ {}& 
            - [\,\boldsymbol\kappa-\tfrac12\, \mathbb{K}\,]\,  
            \eth\, \widehat{\mathbb{N}} 	    
        + \widehat{\mathbb{N}} \left[\,\tfrac12\,\mathbb{K}^{-1}
         \,\eth \boldsymbol\kappa_0 
        + \widehat{\mathbb{K}} \,\mathbbm{k} 
        -q^i \dot{\widehat n}{}^l\,\interior{\rm\bf K}_{li}  
        + q^i\widehat D^l \interior{\rm\bf K}{}_{li}  \right] \, ,  \nonumber     
\end{align}
where, , in virtue of  \eqref{hatextcurv_sq}
\begin{align}\label{hatextcurv_sq0}
\interior{\rm\bf K}{}_{ij} \widehat K^{ij}   = {}& \tfrac14\, {\mathbbm{d}}^{-2} 
\left[ \,2\, \interior{\mathbb{K}}{}_{q\overline{q}}\left( [\,{\mathbbm{a}}^2
+{\mathbbm{b}}\,\overline {\mathbbm{b}})\,]\,{\widehat{\mathbb{K}}{}_{q\overline{q}} } 
- {\mathbbm{a}}\,[\,\overline{\mathbbm{b}}\,{\widehat{\mathbb{K}}{}_{qq}}  
+ {\mathbbm{b}} {\,\overline{\widehat{\mathbb{K}}{}_{qq}} } \,]\,\right) \right.  \nonumber \\ 
{}& \left.\hskip1.2cm + \left\{ \,{\overline{\interior{\mathbb{K}}{}_{qq}}} 
\,[\, {\mathbbm{a}}^2\,{\widehat{\mathbb{K}}{}_{qq}}  
+ {\mathbbm{b}}^2 {\,\overline{\widehat{\mathbb{K}}{}_{qq}} } 
-2\,\mathbbm{a} \,\mathbbm{b} \,\widehat{\mathbb{K}}{}_{q\overline{q}} \,]
+ ``\,CC\," \right\} \right] 
\end{align}
whereas,
$q^i \dot{\widehat n}{}^l\,\interior{\rm\bf K}_{li}$
and $q^i\widehat D^l \interior{\rm\bf K}{}_{li}$ have
been explicitly given in \eqref{5.8}--\eqref{eq:divKnull}.

\section{Future prospects}

We have presented the computational infrastructure necessary for a
numerical code to solve the algebraic-hyperbolic or parabolic-hyperbolic
versions of the constraint equations. The derivatives tangential to the foliation
can be approximated by a finite difference or pseudo-spectral representation of
the $\eth$ or $\boldsymbol{\partial}$ operators.
The ``radial'' integrations along the $\rho$-streamlines can then approximated on a finite grid
by a coupled system of ordinary differential equations by applying the method of lines.
Although the analytic theory
shows that the constraint systems are well-posed under appropriate
conditions, the issue
of a global solution to a nonlinear problem is normally best explored by numerical techniques.
(A  relevant exception is the possibility to apply
energy methods to prove global existence of solutions to both evolutionary versions of the
constraints for initial data near Schwarzschild \cite{phil_i}).

The numerical investigations carried out so far provide some promise
that these evolutionary methods are viable alternatives to the elliptic approach
for the construction of initial data. In an investigation of whether the
algebraic-hyperbolic system
is consistent with asymptotic flatness, 
a code developed in \cite{beyer2} using a spectral representation
of the spin-weighted harmonics \cite{beyer} has
simulated stable evolutions  in the outward $\rho$ direction
for nonlinear perturbations
of Schwarzschild data.
Work in progress in \cite{babac,maciej} using finite difference
codes has shown that the inward
evolution of nonlinear Schwarzschild perturbations can
be stably extended to the interior of the horizon. 
The parabolic-hyperbolic method, combined with a foliation by
spherical surfaces, has been
successfully applied in computing nonlinear perturbations
of Minkowski initial data and using
that data to carry out a corresponding constrained time evolution \cite{Christian}.
For the single black hole case, it has been demonstrated
in \cite{ALI} that the parabolic-hyperbolic system
can be stably and accurately integrated numerically. A detailed investigation of
generic single but boosted and spinning black holes verified that the full
parameter space could be effectively explored
without the use of boundary conditions in the strong field regime.

The ultimate utility of this new approach would be its extension
to multiple black holes. A major concern in such a scheme is the effect of
caustics, where  the ingoing
$\rho$-streamlines focus, or a cross-over surface
$\mycal{S}_X$ where those streamlines
from opposing points of the outer boundary meet.
For a single black hole,
the $\rho$ streamlines can be chosen so that any
caustics and crossovers are inside the apparent horizon,
where the interior can be excised. The excision
of some interior singularity seems to be a necessity
for the application to a spherical foliation \cite{jeff2}.

Formally,  these methods can be applied to the multiple black
hole problem using for the freely specified variables, say, 
a modification of the
superimposed Kerr-Schild data proposed
in~\cite{ksm1,ksm2,i_jeff, racz_bbh,racz_ADM}.
Among other things, the success of a numerical implementation
would depend upon a judicious choice of the foliation $\mycal{S}_\rho$
and the $\rho$-streamlines along which the evolution proceeds.
This is akin to choosing the lapse and shift for
a timelike Cauchy evolution.
For
binary black hole data, although the caustics can be arranged to lie
inside the black holes, the crossover surface  will in
general span the region between them. In that case, unless $\mycal{S}_X$
can be chosen to be a surface of reflection symmetry, as
in the case of data for binary black holes with parallel
or anti-parallel spins the inward evolution 
can produce a discontinuity on $\mycal{S}_X$.
It is anticipated that the methods developed
in \cite{racz_bbh} will be helpful in computing initial data
for binary systems with generic spins and velocities.

Considerable numerical experimentation will be necessary to deal with the technical issues.
The understanding of the analytic
properties and numerical implementation
of the elliptic formulation of the constrains
has had a long and complicated history.
Unlike the iterative global nature of elliptic solvers,
the flexibility of hyperbolic systems
to proceed locally may be of advantage here.
Since hyperbolic evolution of the constraint system can also proceed
in the outward  $\rho$-direction, discontinuities on the crossover surface
$\mycal{S}_X$ can possibly be smoothed and the resulting
data then propagated out to the outer boundary.
 
If  such numerical studies were indeed successful they would open
a new approach to the
question of utmost physical importance: Does the resulting binary black hole
initial data suppress junk radiation?
The sole data needed
on a single large surface in the asymptotic region surrounding the system
distinguishes this approach from other solutions to the constraint problem which rely
on elliptic equations. Whether this feature improves the physical content and control
of the initial data is again a matter for numerical investigation.

\medskip

%%%%%%%%%%%%%%%%%%%% ACKNOWLEDGMENTS %%%%%

\section*{Acknowledgments}

IR and JW were supported in part by the NKFIH grant K-115434 and by NSF grant PHY-1505965 to the University of Pittsburgh, respectively. 

%%%%%%%%%%%%%%%%%%%% FROM HERE COME THE APPENDICES %%%%%%%%

\appendix

\section*{Appendix}
\label{Appendix}
\renewcommand{\theequation}{A.\arabic{equation}}
\renewcommand{\thelemma}{A.\arabic{lemma}}
\renewcommand{\thesubsection}{A.\arabic{subsection}}
\setcounter{equation}{0}

Here we give details of the spin-weight decomposition
of some additional terms that are essential for the implementation
of a numerical code.

\subsection*{Terms involving the lapse $\widehat{N}$}

Using the notation $\widehat{\mathbb{N}}={\widehat N}$,
we obtain
\begin{align}
{}& \widehat D^l\widehat D_l {\widehat N} 
= \widehat \gamma^{kl} [\,\widehat D_k\mathbb{D}_l {\widehat N}\,]  
= \widehat \gamma^{kl} [\,\mathbb{D}_k\mathbb{D}_l {\widehat N}
 - C^{f}{}_{kl} \mathbb{D}_f {\widehat N}\,] 
\\ {}&  \phantom{\widehat D^l\widehat D_l {\widehat N}}  
=  \mathbbm{d}^{-1}\left\{\mathbbm{a}\, q^{kl}-\tfrac12\left[ \mathbbm{b} \,\overline q^k \,\overline q^l 
+ \,\overline{\mathbbm{b}}\, q^k q^l \right]\right\} [\,\mathbb{D}_k\mathbb{D}_l {\widehat N} 
- \tfrac12\,C^{f}{}_{kl}\left[q_f\,\overline q^e
+\,\overline q_f q^e\right] \mathbb{D}_e {\widehat N}\,] \nonumber  \\ {}&  
= \tfrac12\,{\mathbbm{d}}^{-1}[\,\mathbbm{a} \{\,(\eth\,\overline{\eth}\,\widehat{\mathbb{N}}) 
- \mathbb{B}\,(\,\overline{\eth}\,\widehat{\mathbb{N}}) \,\}   - \mathbbm{b} \,\{\,(\,\overline{\eth}^2\,\widehat{\mathbb{N}}) 
- \tfrac12\,\overline{\mathbb{A}}\,(\,\overline{\eth}\,\widehat{\mathbb{N}}) 
- \tfrac12\,\overline{\mathbb{C}} \, ({\eth}\,\widehat{\mathbb{N}}) \,\} + ``\,CC\,"  \,] \,. \nonumber
\end{align}
By virtue of the relation $\dot{\widehat n}{}_k={\widehat n}{}^lD_l{\widehat n}{}_k
=-{\widehat D}_k(\ln{\widehat N})$ we also have 
\begin{equation}
q^{i\,} \dot{\widehat n}{}_i = -  \widehat{\mathbb{N}}^{-1}  \eth \widehat{\mathbb{N}}
\end{equation}
and 
\begin{equation}\label{ndotk}
\mathbbm{k}^{i\,} \dot{\widehat n}{}_i = - (2\,\mathbbm{d} \,\widehat{\mathbb{N}})^{-1}   \{
\,(\eth\,\widehat{\mathbb{N}}) \,[\, \mathbbm{a}\,\overline {\mathbbm{k} }-
\overline {\mathbbm{b}} \, \mathbbm{k} \,]
+ ``\,CC\,"  \, \}\,.
\end{equation}

\subsection*{Terms involving the shift $\widehat N^i$ and ${\rm\bf K}^l{}_{l}$}

By making use of the relations 
\begin{equation}
\widetilde{\mathbb{N}}=q_i\widehat N^i= q_i \widehat\gamma{}^{ij}  \widehat N_j 
= \mathbbm{d}^{-1} (\mathbbm{a}\,q^j - \mathbbm{b}\,\overline q^j)\,{\widehat N}{}_{j}
=  \mathbbm{d}^{-1} (\mathbbm{a}\,\mathbb{N} - \mathbbm{b}\,\overline{\mathbb{N}})%\,,
\end{equation}
or alternatively 
\begin{equation}
\mathbb{N}=q^l {\widehat N}{}_{l}=q^l \widehat\gamma{}_{lk} {\widehat N}{}^{k} = (\mathbbm{a}\,q_k 
+ \mathbbm{b}\,\overline q_k)\,{\widehat N}{}^{k}
= \mathbbm{a}\,\widetilde{\mathbb{N}} + \mathbbm{b}\,\overline {\widetilde{\mathbb{N}}} \,, 
\end{equation}
the Lie derivative $\mycal{L}_{\widehat n}\,({\rm\bf K}^l{}_{l})$ appearing in (\ref{constr_mom2})
can be expressed as 
\begin{align}\label{Lie_K} 
\mycal{L}_{\widehat n}\,({\rm\bf K}^l{}_{l}) 
= {}& {\widehat n}{}^i D_i{\rm\bf K}^l{}_{l} 
= {{\widehat N}^{-1}} [\, (\partial_\rho)^i -\widehat N^i \,]\, D_i {\rm\bf K}^l{}_{l} 
= {\widehat N}^{-1} [\, \partial_\rho{\rm\bf K}^l{}_{l}  
-\widehat N^i \, \mathbb{D}_i{\rm\bf K}^l{}_{l} \,]\nonumber  \\ 
= {}&  \mycal{L}_{\widehat n}\,\mathbb{K} = \,\widehat{\mathbb{N}}^{-1}[\, (\partial_\rho \mathbb{K})
-\tfrac12\, \widetilde{\mathbb{N}} \,(\,\overline{\eth}\, \mathbb{K}) 
- \tfrac12\,\overline{\widetilde{\mathbb{N}}}\, (\eth\,\mathbb{K})    \,]\,,
\end{align}
where
\begin{equation}
\mathbb{K} = {\rm\bf K}^l{}_{l}= \widehat\gamma^{kl} \,{\rm\bf K}{}_{kl}
\end{equation} 
and we have used  $\widehat N^i D_i{\rm\bf K}^l{}_{l} = \widehat N^i\mathbb{D}_i{\rm\bf K}^l{}_{l}
=\tfrac12\,\widehat N^i\left(q_i\,\overline q^j+\,\overline q_iq^j\right) \mathbb{D}_j{\rm\bf K}^l{}_{l}$\, .

\subsection*{Terms involving the trace-free part of ${\rm\bf K}{}_{kl}$}

By setting 
\begin{equation}
\interior{\mathbb{K}}{}_{qq} = q^kq^l\,\interior{\rm\bf K}{}_{kl}
\end{equation}
and
\begin{equation}
\interior{\mathbb{K}}{}_{q\overline{q}} = q^{k}\,\overline q^l\,\interior{\rm\bf K}{}_{kl}\,,
\end{equation}
in virtue of  (\ref{intK}), we obtain
\begin{equation}\label{decomp_bfK}
\interior{\rm\bf K}{}_{ij}
 =\tfrac12\,q{}_{ij}\,\interior{\mathbb{K}}{}_{q\overline{q}}+\tfrac14\,[\, q_{i}q_j\,\overline{\interior{\mathbb{K}}{}_{qq}}
+ \overline q_{i}\overline q_j\,\interior{\mathbb{K}}{}_{qq}\,] \,.
\end{equation}

\medskip

Note that, since $\interior{\rm\bf K}{}_{kl}$ is trace-free,
 $\interior{\mathbb{K}}{}_{q\overline{q}}$
and $\interior{\mathbb{K}}{}_{qq}$ are not functionally independent.
Indeed, the trace-free condition $\widehat\gamma^{kl}\,\interior{\rm\bf K}{}_{kl}=0$ implies 
\begin{equation}\label{relation}
\interior{\mathbb{K}}{}_{q\overline{q}}
= (2\,\mathbbm{a})^{-1} [\,\mathbbm{b}\,\overline{\interior{\mathbb{K}}{}_{qq}} 
+  \overline{\mathbbm{b}}\,\interior{\mathbb{K}}{}_{qq} \,]\,.
\end{equation}

For both $\mathbbm{a}^{-1}$ and $\interior{\mathbb{K}}{}_{q\overline{q}}$, to be well-defined
$\mathbbm{a}$ cannot vanish.
This is guaranteed because $\widehat{\gamma}_{ij}$ is a positive definite
Riemannian metric so that $\mathbbm{d}=\mathbbm{a}^2-\mathbbm{b}\,\overline{\mathbbm{b}}$
must be positive. 

\medskip

We then have
\begin{equation}\label{qndotK}
q^{i\,} \dot{\widehat n}{}^{k\,} \interior{\rm\bf K}{}_{ki}	 =
-\tfrac12 ( {\widehat{\mathbb{N}}}\,{\mathbbm{d}})^{-1} \left[
\mathbbm{a}\,(\overline \eth\, \widehat{\mathbb{N}} ) \, \interior{\mathbb{K}}{}_{qq}    
+  \mathbbm{a}\,( \eth {\widehat{\mathbb{N}}} ) \, \interior{\mathbb{K}}{}_{q\overline{q}}  
-\mathbbm{b}\,(\overline \eth {\widehat{\mathbb{N}}} ) \, \interior{\mathbb{K}}{}_{q\overline{q}} 
- \overline{\mathbbm{b}}\,(\eth {\widehat{\mathbb{N}}} ) \,
\interior{\mathbb{K}}{}_{qq}   \right] ,
\end{equation}
\begin{align}\label{qDK}
q^i \widehat D^{k\,} \interior{\rm\bf K}{}_{ki}	 &=
\tfrac12\,{\mathbbm{d}}^{-1} \left\{
\mathbbm{a}\,\overline \eth \, \interior{\mathbb{K}}{}_{qq}    
+  \mathbbm{a}\,\eth \, \interior{\mathbb{K}}{}_{q\overline{q}}  
-\mathbbm{b} \,\overline \eth \, \interior{\mathbb{K}}{}_{q\overline{q}} 
- \overline{\mathbbm{b}}\, \eth \, \interior{\mathbb{K}}{}_{qq}   \right\}
\nonumber \\
& - \frac{ \mathbbm{a}} {4\mathbbm{d}} \left\{
3\,\overline{\mathbbm{B}} \,\interior{\mathbb{K}}{}_{qq} 
+3\,{\mathbbm{B}}\, \interior{\mathbb{K}}{}_{q\overline{q}} 
+{\mathbbm{A}} \,\interior{\mathbb{K}}{}_{q\overline{q}} +{\mathbbm{C}}\,
\overline{\interior{\mathbb{K}}{}_{qq}} \right\} 
\nonumber \\
& + \frac{ \mathbbm{b}} {4\mathbbm{d}} \left\{
\overline{\mathbbm{C}} \,\interior{\mathbb{K}}{}_{qq} 
+\overline{\mathbbm{A}} \,\interior{\mathbb{K}}{}_{q\overline{q}} 
+\overline{\mathbbm{B}} \,\interior{\mathbb{K}}{}_{q\overline{q}} +{\mathbbm{B}} \,\overline{\interior{\mathbb{K}}{}_{qq}} \right\} 
% \nonumber \\ &
+ \frac{\overline{ \mathbbm{b}}} {2\mathbbm{d}} \left\{
\mathbbm{A}\,\interior{\mathbb{K}}{}_{qq} 
 +\mathbbm{C}\, \interior{\mathbb{K}}{}_{q\overline{q}}  \right\}  ,
\end{align}
\begin{align}\label{tracefreebfcurv_sq}
\interior{\rm\bf K}{}_{ij}   \interior{\rm\bf K}{}^{ij}  
= {} & \tfrac14  \,{\mathbbm{d}}^{-2} \left[
\left \{ \, {\overline { \interior{\mathbb{K}}{}_{qq}}} \,
(\, {\mathbbm{a}}^2\,{ \interior{\mathbb{K}}{}_{qq}}   
+ {\mathbbm{b}}^2  {\,\overline { \interior{\mathbb{K}}{}_{qq}}}  
-4\,\mathbbm{a} \,\mathbbm{b} \,\interior{\mathbb{K}}{}_{q\overline{q}}\, )
+ ``\,CC\," \right \} % \\ {} & 
+  2\,  ({\mathbbm{a}}^2+{\mathbbm{b}}\,\overline {\mathbbm{b}} )
\,\interior{\mathbb{K}}{}_{q\overline{q}}^2\right] \,.
% \nonumber
\end{align}

\subsection*{The determination of $\mycal{L}_{\widehat n}\,{\rm\bf k}{}_{l}$ and
$\mycal{L}_{\rho} {\rm\bf k}{}_{i}$}

The Lie derivative $\mycal{L}_{\widehat n}\,{\rm\bf k}{}_{l}$ appearing in
(\ref{constr_mom1}), can be re-expressed as follows.
First, note that  
\begin{equation}
\left(\mycal{L}_{\widehat n}\,{\rm\bf k}{}_{l}\right)  {\widehat n}{}^l
= \mycal{L}_{\widehat n}\left({\rm\bf k}{}_{l}\,{\widehat n}{}^l\right) =0 ,
\end{equation}
which implies
\begin{equation}
\mycal{L}_{\widehat n}\,{\rm\bf k}{}_{l}
= \widehat\gamma_{\,l}{}^i \mycal{L}_{\widehat n}\,{\rm\bf k}{}_{i}\,.
\end{equation}
Then, it is straightforward to verify that 
\begin{align}
\mycal{L}_{\widehat n}\,{\rm\bf k}{}_{l} 
= {}& \widehat\gamma_{\,l}{}^i \mycal{L}_{\widehat n}\,{\rm\bf k}{}_{i}
={{\widehat N}^{-1}} \widehat\gamma_{\,l}{}^i\,[\, \mycal{L}_{\rho} {\rm\bf k}{}_{i} 
- \mycal{L}_{\widehat N} {\rm\bf k}{}_{i} \,] \nonumber \\ = {}&  
{{\widehat N}^{-1}} [\, \widehat\gamma_{\,l}{}^i(\mycal{L}_{\rho} {\rm\bf k}{}_{i}) 
 - \widehat N^f \widehat D_f {\rm\bf k}{}_{l}
- {\rm\bf k}{}_{f} \widehat D_l \widehat N^f  \,] \nonumber \\ = {}&  
{{\widehat N}^{-1}} [\, \widehat\gamma_{\,l}{}^i(\mycal{L}_{\rho} {\rm\bf k}{}_{i}) 
 - \widehat N^f \mathbb D_f {\rm\bf k}{}_{l} 
- {\rm\bf k}{}_{f} \mathbb D_l \widehat N^f  \,] \,,
\end{align}
where in the second line we have used the torsion free property of the
connection when evaluating $\mycal{L}_{\widehat N} {\rm\bf k}{}_{i}$ .

In determining $q^l \mycal{L}_{\widehat n}\,{\rm\bf k}{}_{l} $ we use 
\begin{equation}
q^l\widehat\gamma_{\,l}{}^i(\mycal{L}_{\rho} {\rm\bf k}{}_{i})
=q^l q_{\,l}{}^i(\mycal{L}_{\rho} {\rm\bf k}{}_{i}) 
= (\partial_\rho \mathbbm{k})%\,,
\end{equation}
and
\begin{align}
q^l\,[\,\widehat N^f \mathbb D_f {\rm\bf k}{}_{l}\, + {\rm\bf k}{}_{f} \mathbb D_l \widehat N^f ] = 
\tfrac{1}{2}[  \, {\widetilde{\mathbb{N}}} \, \,\overline{\eth}\,\mathbbm{k} 
+ \,\overline {\widetilde{\mathbb{N}}} \, \eth\,\mathbbm{k}  \, ] 
+\tfrac{1}{2} [ \, \mathbbm{k} \, \eth \,\overline {\widetilde{\mathbb{N}}} 
+ \,\overline {\mathbbm{k}} \, \eth {\widetilde{\mathbb{N}}}  \,  ] .
\end{align}
Then
\begin{equation}\label{liek}
q^l \mycal{L}_{\widehat n}\,{\rm\bf k}{}_{l} ={\widehat{\mathbb{N}}^{-1}} \left( \partial_\rho \mathbbm{k}
- \tfrac12 [ \, {\widetilde{\mathbb{N}}} \, \,\overline{\eth}\,\mathbbm{k} 
+ \,\overline {\widetilde{\mathbb{N}}} \, \eth\,\mathbbm{k}  
+ \mathbbm{k} \, \eth \,\overline {\widetilde{\mathbb{N}}} 
+ \,\overline {\mathbbm{k}} \, \eth \,{\widetilde{\mathbb{N}}}  \,  ] \right) .
\end{equation}

\subsection*{The decomposition of $\widehat D_{k} {\widehat N}{}_{l}$}

We also need to evaluate the auxiliary expressions 
$q^kq^l\,(\widehat D_{k} {\widehat N}{}_{l})$ and $\,\overline q^kq^l\,(\widehat D_{k} {\widehat N}{}_{l})$. 
To do so, first notice that 
\begin{equation}
\widehat D_{k} {\widehat N}{}_{l}={\mathbb D}_{k} {\widehat N}{}_{l} 
-{C^f}{}_{kl}  {\widehat N}{}_{f} \,,
\end{equation}
from which 
\begin{align}
q^kq^l\,(\widehat D_{k} {\widehat N}{}_{l}) = {}& q^kq^l\,({\mathbb D}_{k} {\widehat N}{}_{l}) 
- q^kq^l\,{C^f}{}_{kl}\,[ \tfrac12\,(q_f\,\overline q^e + \,\overline q_fq^e) ]  {\widehat N}{}_{e} \nonumber \\ 
= {}&  \eth\,\mathbb{N} -\tfrac12\, \mathbb{C}\,\overline{\mathbb{N}} 
  - \tfrac12\,\mathbb{A}\,\mathbb{N}\,,
\end{align}
\begin{align}
\,\overline q^kq^l\,(\widehat D_{k} {\widehat N}{}_{l}) 
= {}& \,\overline q^kq^l\,({\mathbb D}_{k} {\widehat N}{}_{l}) 
- \,\overline q^kq^l\,{C^f}{}_{kl}\,[ \tfrac12\,(q_f\,\overline q^e 
+ \,\overline q_f\,q^e) ]  {\widehat N}{}_{e} \nonumber \\ 
= {}&  \overline\eth\,\mathbb{N} -\tfrac12\, \mathbb{B}\,\overline{\mathbb{N}} 
-\tfrac12\,\overline{\mathbb{B}}\,\mathbb{N} \,.
\end{align}

\subsection*{Terms involving $\widehat K_{ij}$}

Before determining $q^l\,[\,\widehat\gamma{}^{ef} {\rm\bf k}{}_{e} \widehat K_{fl}\,]$,
we need to evaluate 
the extrinsic curvature $\widehat K_{ij}$ of $\mycal{S}_\rho$
as given by (\ref{hatextcurv}),
\begin{align}\label{hatextcurv2}
\widehat K_{ij}= {}& \tfrac12\,\mycal{L}_{\widehat n} {\widehat \gamma}_{ij}
=\tfrac12\,{\widehat N}^{-1}[\,\mycal{L}_{\rho}{\widehat \gamma}_{ij}
- ( \widehat D_i\widehat N_j + \widehat D_j\widehat N_i )] \\ 
= {}& \tfrac12\,{\,\widehat{\mathbb{N}}}^{-1}[(\partial_\rho\mathbbm{a})\,q_{ij}
+\tfrac12\,[\left(\partial_\rho\mathbbm{b}\right)\,\overline q_i\,\overline q_j 
+ \left(\partial_\rho\,\overline{\mathbbm{b}}\right) q_i q_j] 
- (\widehat D_i\widehat N_j + \widehat D_j\widehat N_i )]\,, \nonumber
\end{align}
where in the last step (\ref{lie_dragged}) was applied.  As a result, 
\begin{align}\label{hatextcurv_trace}
{}& \,\widehat{\mathbb{K}} = \widehat K{}^{\,l}{}_{l} 
=   \widehat \gamma^{ij} \widehat K_{ij}
= \mathbbm{d}^{-1}\left\{\mathbbm{a}\, q^{ij}
-\tfrac12\left[ \mathbbm{b} \,\overline q^i \,\overline q^j
+ \,\overline{\mathbbm{b}}\, q^i q^j \right]\right\}\widehat K_{ij} 
= \tfrac12\,({\,\widehat{\mathbb{N}}\,\mathbbm{d}})^{-1} \times \nonumber  \\  
{}& \times \bigl[\mathbbm{a} \{(\partial_\rho\mathbbm{a}) 
- q^i \,\overline q^j\,[\,\widehat D_i\widehat N_j + \widehat D_j\widehat N_i\,] \} 
- \mathbbm{b}\{(\partial_\rho\,\overline{\mathbbm{b}}) 
- \,\overline q^i \,\overline q^j (\widehat D_i\widehat N_j)\}  \bigr] + ``\,CC\,"  \nonumber  \\ 
{}& = \tfrac12\,({\,\widehat{\mathbb{N}}\,\mathbbm{d}})^{-1}
\left\{\mathbbm{a}\,[\,(\partial_\rho\mathbbm{a}) - (\,\overline\eth\,\mathbb{N}) 
+ \,\overline{\mathbb{B}}\,\mathbb{N} \,] \right. \nonumber 
\\ {}& \hskip2.5cm \left.  - \mathbbm{b}\,[\,(\partial_\rho\,\overline{\mathbbm{b}})
 - \,(\,\overline{\eth}\,\overline{\mathbb{N}}) 
+\tfrac12\, \,\overline{\mathbb{C}}\,\mathbb{N} 
+\tfrac12\, \,\overline{\mathbb{A}}\,\overline{\mathbb{N}} \,]  \right\} + ``\,CC\," \, .
\end{align}

Now set
\begin{equation}\label{hatextcurv_qq}
\,\widehat{\mathbb{K}}{}_{qq} =  q^i q^j\widehat K_{ij} 
=\tfrac12\,{\,\widehat{\mathbb{N}}}^{-1}\left\{2\,\partial_\rho\mathbbm{b} 
- 2\,\eth\,\mathbb{N}
+ {\mathbb{C}}\,\overline {\mathbb{N}} +\mathbb{A}\, {\mathbb{N}} \,  \right\} \,, 
\end{equation}
\begin{equation}\label{hatextcurv_qbq}
\,\widehat{\mathbb{K}}{}_{q\overline{q}} =  q^i \,\overline q^j\widehat K_{ij} 
=\tfrac12\,{\,\widehat{\mathbb{N}}}^{-1}\left\{2\,\partial_\rho\mathbbm{a}
- \,\overline \eth\,\mathbb{N}  - \eth\,\overline{\mathbb{N}}
+ {\mathbb{B}}\,\overline {\mathbb{N}} +\,\overline{\mathbb{B}}\,{\mathbb{N}} \,  \right\} \,. 
\end{equation}
Then, because the symmetric 2-tensor $\widehat K{}^{\,l}{}_{l}$ is determined
by three real functions,
it follows that $\widehat{\mathbb{K}}{}_{q\overline{q}}$, $\widehat{\mathbb{K}}{}_{qq}$
and $\widehat{\mathbb{K}}$
are functionally dependent. In determining their algebraic relation
we introduce the auxiliary variables
\begin{equation}\label{hatextcurv_qbq0}
{}^\star\hskip-.051mm \widehat{\mathbb{K}}{}_{q\overline{q}}
 =  q^i \,\overline q^j\,[ \widehat K_{ij} 
 -\tfrac12\,\widehat \gamma_{ij} \widehat K{}^{\,l}{}_{l} ]
= \widehat{\mathbb{K}}{}_{q\overline{q}} - {\mathbbm{a}} \,\widehat{\mathbb{K}} \, , 
\end{equation}
\begin{equation}\label{hatextcurv_qbq1}
{}^\star\hskip-.051mm \widehat{\mathbb{K}}{}_{qq} 
=  q^i \, q^j\,[ \widehat K_{ij} -\tfrac12\,\widehat \gamma_{ij} \widehat K{}^{\,l}{}_{l} ]
= \widehat{\mathbb{K}}{}_{qq} - {\mathbbm{b}} \,\widehat{\mathbb{K}}  \,.
\end{equation}
The analog of the trace relation (\ref{relation}) then gives
\begin{equation}\label{relation1}
{}^\star\hskip-.051mm \widehat{\mathbb{K}}{}_{q\overline{q}}
= (2\,\mathbbm{a})^{-1} [\,\mathbbm{b}\,\overline{{}^\star
\hskip-.051mm \widehat{\mathbb{K}}{}_{qq}} 
+  \overline{\mathbbm{b}}\,{}^\star\hskip-.051mm \widehat{\mathbb{K}}{}_{qq}  \,]\,,
\end{equation}
from which it follows, in virtue of (\ref{hatextcurv_qbq0}) and (\ref{hatextcurv_qbq1}), 
\begin{equation}\label{relation2}
\widehat{\mathbb{K}}{}_{q\overline{q}}
={\mathbbm{a}}^{-1}\{\, \mathbbm{d}\cdot\widehat{\mathbb{K}} 
+ \tfrac12 \,[\,\mathbbm{b}\,\overline{\widehat{\mathbb{K}}{}_{qq}} 
+  \overline{\mathbbm{b}}\,\widehat{\mathbb{K}}{}_{qq}  \,]\,\}\, .
\end{equation}

Then, by making use of all the prior variables related to $\widehat K^{ij}$, we obtain
\begin{equation}\label{hatextcurv_uqq}
q_i q_j \widehat K^{ij} 
={\mathbbm{d}}^{-2}\,[\, {\mathbbm{a}}^2\,{\widehat{\mathbb{K}}{}_{qq}}  
+ {\mathbbm{b}}^2 {\,\overline{\,\widehat{\mathbb{K}}{}_{qq}} } 
-2\,\mathbbm{a}\, \mathbbm{b} \,\widehat{\mathbb{K}}{}_{q\overline{q}} \,  ] %\,, 
\end{equation}
and
\begin{equation}\label{hatextcurv_uqbq}
q_i \,\overline q_j \widehat K^{ij} ={\mathbbm{d}}^{-2}[\,({\mathbbm{a}}^2
+{\mathbbm{b}}\,\overline {\mathbbm{b}} )\,{\widehat{\mathbb{K}}{}_{q\overline{q}} } 
-\mathbbm{a} \,\overline{\mathbbm{b}} \,\widehat{\mathbb{K}}{}_{qq} 
-\mathbbm{a} \,\mathbbm{b} \, {\,\overline{\widehat{\mathbb{K}}{}_{qq}}} \,  ] \,.
\end{equation}
These relations, along with (\ref{decomp_bfK}), imply
\begin{align}\label{hatextcurv_sq}
\interior{\rm\bf K}{}_{ij} \widehat K^{ij}   = {}& \tfrac14\, {\mathbbm{d}}^{-2} 
\left[ \,2\, \interior{\mathbb{K}}{}_{q\overline{q}}\left( [\,{\mathbbm{a}}^2
+{\mathbbm{b}}\,\overline {\mathbbm{b}})\,]\,{\widehat{\mathbb{K}}{}_{q\overline{q}} } 
- {\mathbbm{a}}\,[\,\overline{\mathbbm{b}}\,{\widehat{\mathbb{K}}{}_{qq}}  
+ {\mathbbm{b}} {\,\overline{\widehat{\mathbb{K}}{}_{qq}} } \,]\,\right) \right.  \nonumber \\ 
{}& \left.\hskip1.2cm + \left\{ \,{\overline{\interior{\mathbb{K}}{}_{qq}}} 
\,[\, {\mathbbm{a}}^2\,{\widehat{\mathbb{K}}{}_{qq}}  
+ {\mathbbm{b}}^2 {\,\overline{\widehat{\mathbb{K}}{}_{qq}} } 
-2\,\mathbbm{a} \,\mathbbm{b} \,\widehat{\mathbb{K}}{}_{q\overline{q}} \,]
+ ``\,CC\," \right\} \right] 
\end{align}
and
\begin{align}
\widehat K_{ij} \widehat K^{ij}   = {}& \tfrac14\, {\mathbbm{d}}^{-2} 
\left \{ \,\overline{\,\widehat{\mathbb{K}}{}_{qq}}
\left [ {\mathbbm{a}}^2\,\widehat{\mathbb{K}}{}_{qq} 
+ {\mathbbm{b}}^2 \,\overline{\,\widehat{\mathbb{K}}{}_{qq} } 
-4\,\mathbbm{a} \,\mathbbm{b} \,\widehat{\mathbb{K}}{}_{q\overline{q}} \right ]
+ ``\,CC\," \right \} % \\{}& 
+  \tfrac12\, {\mathbbm{d}}^{-2} 
({\mathbbm{a}}^2+{\mathbbm{b}}\,\overline {\mathbbm{b}} )
\,\widehat{\mathbb{K}}{}_{q\overline{q}}^2  . \label{key} 
\end{align}

Finally, the spin-weighted analogue of (\ref{Lie_K}) is  
\begin{align}\label{Lie_hatK} 
\mycal{L}_{\widehat n}\,({\widehat K}^l{}_{l}) 
= {}& \mycal{L}_{\widehat n}\,{\,\widehat{\mathbb{K}}} 
=\,\widehat{\mathbb{N}}^{-1}\left[ (\partial_\rho \,\widehat{\mathbb{K}} )
 -\tfrac12\, \widetilde{\mathbb{N}} \,(\,\overline{\eth}\, \,\widehat{\mathbb{K}})
- \tfrac12\,\overline{\widetilde{\mathbb{N}}}\, (\eth \,\widehat{\mathbb{K}})    \right]\,.
\end{align}

\subsection*{Terms involving starred quantities}

By virtue of \eqref{hatextcurv} and \eqref{instarK}, 
\begin{equation}
\instar{{K}}_{ij}= \widehat{N}\,\widehat{K}_{ij} \, .
\end{equation}
Accordingly, as a consequence of \eqref{hatextcurv2}--\eqref{key}, 
\begin{align}\label{starextcurv_trace}
{}& \,\instar{\mathbb{K}} 
=  \tfrac12\,\mathbbm{d}^{-1}\left\{\mathbbm{a}\,[\,(\partial_\rho\mathbbm{a}) 
- (\,\overline\eth\,\mathbb{N}) 
+ \,\overline{\mathbb{B}}\,\mathbb{N} \,] \right. \nonumber 
\\ {}& \hskip2.5cm \left.  - \mathbbm{b}\,[\,(\partial_\rho\,\overline{\mathbbm{b}})
 - \,(\,\overline{\eth}\,\overline{\mathbb{N}}) +\tfrac12\, \,\overline{\mathbb{C}}\,\mathbb{N} 
+\tfrac12\, \,\overline{\mathbb{A}}\,\overline{\mathbb{N}} \,]  \right\} + ``\,CC\," \, .
\end{align}
By now setting 
\begin{equation}\label{starextcurv_qq}
\,\instar{\mathbb{K}}{}_{qq} =  q^i q^j\instar K_{ij} 
=\tfrac12\,\left\{2\,\partial_\rho\mathbbm{b} - 2\,\eth\,\mathbb{N}
+ {\mathbb{C}}\,\overline {\mathbb{N}} +\mathbb{A}\, {\mathbb{N}} \,  \right\} \,, 
\end{equation}
\begin{equation}\label{starextcurv_qbq}
\,\instar{\mathbb{K}}{}_{q\overline{q}} =  q^i \,\overline q^j\instar K_{ij} 
=\tfrac12\,\left\{2\,\partial_\rho\mathbbm{a}
- \,\overline \eth\,\mathbb{N}  - \eth\,\overline{\mathbb{N}}
+ {\mathbb{B}}\,\overline {\mathbb{N}} +\,\overline{\mathbb{B}}\,{\mathbb{N}} \,  \right\} 
\end{equation}
we obtain
\begin{equation}\label{relation2star}
\instar{\mathbb{K}}{}_{q\overline{q}}
={\mathbbm{a}}^{-1}\{\, \mathbbm{d}\cdot\instar{\mathbb{K}} 
+ \tfrac12 \,[\,\mathbbm{b}\,\overline{\instar{\mathbb{K}}{}_{qq}} 
+  \overline{\mathbbm{b}}\,\instar{\mathbb{K}}{}_{qq}  \,]\,\}\, ,
\end{equation}
while \eqref{5.10} and \eqref{5.11} follow straightforwardly from
\eqref{hatextcurv_sq} and \eqref{key}.

%%%%%%%%%%%%%%%%%%%% REFERENCES 

\end{document}